\documentclass[a4paper,12pt]{article}
\pdfoutput=1
\usepackage{epsfig,graphicx,xcolor,amsbsy,amssymb,latexsym,amsfonts,amsmath,setspace}
\usepackage{pstricks}
\usepackage{color}
\usepackage{soul}
\usepackage[numbers,square,comma, compress]{natbib}
\usepackage{placeins}
\usepackage{wrapfig}

\usepackage{eurosym}
\usepackage[a4paper]{geometry}
\usepackage{graphicx}
\usepackage{tikz}
\usepackage{xcolor}
\usepackage{amsmath,amssymb,amsfonts,pstricks,setspace}
\usepackage[enableskew]{youngtab}

\numberwithin{equation}{section}

\newcommand{\bea}{\begin{eqnarray}\displaystyle}
\newcommand{\eea}{\end{eqnarray}}

\newcommand{\figref}[1]{Fig.~\protect\ref{#1}}

\newcommand{\ct}[2]{\vartheta_{\alpha_{#1}\alpha_{#2}}}

\title{
\begin{flushright}{\vspace{-2.5cm}\small LYCEN 2018-09\\}\end{flushright}
\vspace{2.3cm}
\bf{Five-Dimensional Gauge Theories  \\ from \\
Shifted Web Diagrams}\\[15pt]}
\author{\large \textsc{Brice Bastian\footnote{\tt b.bastian@ipnl.in2p3.fr}},~~\textsc{Stefan~Hohenegger\footnote{\tt s.hohenegger@ipnl.in2p3.fr},~ Amer Iqbal\footnote{\tt  amer@alum.mit.edu},~ Soo-Jong Rey\,\footnote{\tt rey.soojong@gmail.com}}}
\date{}

\begin{document}

\maketitle

\begin{center}
\renewcommand{\thefootnote}{\fnsymbol{footnote}}\vspace{-0.5cm}
${}^{\footnotemark[1]\footnotemark[2]}$ Universit\'e de Lyon\\
UMR 5822, CNRS/IN2P3, Institut de Physique Nucl\'eaire de Lyon\\ 4 rue Enrico Fermi, 69622 Villeurbanne Cedex, \rm FRANCE\\[0.4cm]
${}^{\footnotemark[3]}$ Abdus Salam School of Mathematical Sciences \\ Government College University, Lahore, PAKISTAN\\[0.4cm]
${}^{\footnotemark[3]}$ Center for Theoretical Physics, Lahore, PAKISTAN\\[0.4cm]
${}^{\footnotemark[4]}$ School of Physics and Astronomy, Seoul National University\\
Gwanak-ro 1, Gwanak-gu, Seoul 08826 \rm KOREA\\[1cm]
\end{center}

\begin{abstract}
In previous works (arXiv:1610.07916, arXiv:1711.07921, arXiv:1807.00186) we studied a class of toric Calabi-Yau threefolds which engineer six-dimensional supersymmetric gauge theories with gauge group $U(N)$ and adjoint matter. The K\"ahler moduli space of these manifolds can be extended through flop transformations to include regions which are described by so-called shifted toric web diagrams. In this paper we analyse gauge theories that are engineered by these shifted toric web diagrams and argue that in specific limits, some of the them engineer five-dimensional quiver gauge theories with gauge group $G\subset U(N)$ and with fundamental and bi-fundamental matter. We discuss several examples in detail and describe how the matter sector is obtained from the six-dimensional parent theory.
 \end{abstract}

\newpage

\tableofcontents

\onehalfspacing

\vskip1cm

\section{Introduction}
The engineering of five- and six-dimensional superconformal theories in string theory, using branes or geometry, has provided insights into various properties of these theories including the allowed matter content. More recently, the classification of the five- and six-dimensional superconformal theories has been studied using M- and F-theory on Calabi-Yau threefolds \cite{Heckman:2013pva,Heckman:2015bfa,DelZotto:2017pti,Jefferson:2017ahm,Jefferson:2018irk}. An extension of the classification of six-dimensional superconformal theories also resulted in the classification of little string theories \cite{Bhardwaj:2015oru}. The latter have an intrinsic scale set by $M_{\text{string}}$ and are related to superconformal theories by a particular limit which gets rid of this scale \cite{Seiberg:1996vs,Berkooz:1997cq,Blum:1997mm,Seiberg:1997zk,Losev:1997hx,Intriligator:1997dh,Aharony:1998ub,Aharony:1999ks,Kutasov:2001uf}. The geometry of Calabi-Yau threefolds plays an important role in these classifications determining the gauge groups and matter content of the available theories \cite{Witten:1996qb,Morrison:1996na,Morrison:1996pp,Chou:1997ba}.

In this paper we continue our study of a particularly rich class of such Calabi-Yau threefolds, called $X_{N,M}$ (with $N,M\in\mathbb{N}$), which was introduced in \cite{Haghighat:2013gba,Haghighat:2013tka,Hohenegger:2013ala}. These manifolds are toric \cite{Kan} and are parametrized by $NM+2$ independent K\"ahler parameters. It was proposed in \cite{Hohenegger:2016yuv} that a given $X_{N,M}$ engineers via F-theory three different quiver gauge theories in six dimensions (compactified on $S^1$) with gauge groups $[U(N)]^M$, $[U(M)]^N$ and $[U(NM/k)]^k$ respectively, where $k=\text{gcd}(N,M)$. These theories are dual to each other in the sense that they have the same partition function $\mathcal{Z}_{N,M}(\{\mathbf{h},\mathbf{v},\mathbf{m}\})$, where $\{\mathbf{h},\mathbf{v},\mathbf{m}\}$ is the set of K\"ahler parameters of $X_{N,M}$ (with a suitable parametrisation). The main difference between the three gauge theories is the interpretation of these K\"ahler parameters in terms of the gauge theory data, namely coupling constants, the mass parameters of the matter content as well as parameters related to the gauge structure.

It was further pointed out in \cite{Bastian:2018dfu} that the number of (different) dual gauge theories is in general even higher than three. Indeed, it was argued in \cite{Hohenegger:2016yuv} (and demonstrated explicitly in \cite{Bastian:2017ing} for $\text{gcd}(N,M)=1$) that the Calabi-Yau manifolds $X_{N,M}$ and $X_{N',M'}$ (with $NM=N'M'$ and $\text{gcd}(N,M)=\text{gcd}(N',M')$) are related through a series of flop transformations that leave $\mathcal{Z}_{N,M}(\{\mathbf{h},\mathbf{v},\mathbf{m}\})$ invariant. This implies that $X_{N,M}$ engineers a web of dual gauge theories with gauge groups $[U(N')]^{M'}$ for all $(M',N')$ satisfying the above conditions. 

The duality transformation discovered in \cite{Hohenegger:2016yuv} is particular in the sense that it relates specific classes of toric web diagrams to one another. However, the extended moduli space \cite{Reid,CT1,CT2,CT3,CT4,Witten:1996qb,Chou:1997ba} of $X_{N,M}$ also contains other geometries, such as the so-called \emph{shifted webs} $X_{N,M}^{(\delta)}$, which were introduced in \cite{Bastian:2018dfu}. The latter are parametrised by an integer $\delta\in \{0,1,\ldots,N-1\}$ which governs how the planar web is glued together on a torus (an example for $M=1$ is shown in \figref{Fig:TwistExample}), such that $X_{N,M}^{(\delta=0)}=X_{N,M}$. While for $\delta=0$ F-theory compactifications on $X_{N,M}^{(\delta=0)}$ are dual to M5-brane configurations probing a non-trivial background, a similar formulation for the cases $\delta\neq 0$ is more delicate. In particular, while a straight-forward decompactification limit exists in the former case, a similar limit is not immediately obvious in the latter case. This also makes the question what gauge theories (if any) are engineered by the configurations $X_{N,M}^{(\delta\neq 0)}$ more difficult to answer, since a simple five-dimensional limit (in which the partition function $\mathcal{Z}_{N,M}$ could be compared with the Nekrasov partition function) is not evident. In this paper, we analyze this problem for the case of shifted web diagrams of the form $X_{N,1}^{(\delta)}$. We show first with the help of a number of examples, that such web diagrams allow non-trivial one-parameter limits which reduce them to configurations that are known to engineer five-dimensional quiver gauge theories. While the gauge group of the latter is in general a subgroup of $U(N)$, they generically exhibit a rich matter spectrum. Based on these examples, we can generalize our discussion by extending a recurrent pattern.

Specifically, this paper is organised as follows: In section~\ref{Examples} we study various examples of the decompactification of shifted webs which reduce the six-dimensional theory to a five-dimensional theory. Different decompactification limits give rise to various different five-dimensional theories which can be thought of as dual to one another once lifted back up to six dimensions. We particularly point out that certain well known five-dimensional theories arise via the decompactification of manifolds that are described by the the above mentioned shifted web diagrams. Generalising a pattern appearing in the examples of section~\ref{Examples}, we propose a two parameter series of five-dimensional theories in section~\ref{TwoPara}, which arises from the decompactification limit of shifted web diagrams of the type $X_{A+B+AB,1}^{(B(A+1)-1)}$ for $A,B\in\mathbb{N}$ (subject to certain conditions). In section~\ref{Sect:Conclusions}  we present our conclusions and comment on future directions. Furthermore, several technical details have been relegated to two appendices.
\section{Specific examples}
\label{Examples}
The two-parameter class of Calabi-Yau threefolds (labelled interchangeably $X_{N,M}$ or  $X^{(\delta=0)}_{N,M}$ in the current work) that was studied in \cite{Hohenegger:2016eqy,Hohenegger:2016yuv,Bastian:2017ing,Bastian:2017ary} is dual to $N$ M5-branes wrapped on $S^1$ probing a transverse ALE-space of type $\widehat{A}_{M-1}$. The K\"ahler moduli space of these manifolds takes the form of a cone which can be extended through flop transitions of $(-1,-1)$-curves\footnote{We denote curves with local geometry ${\cal O}(a)\oplus{\cal O}(b)\mapsto \mathbb{P}^1$ as $(a,b)$-curves, where $a+b=-2$ due to the Calabi-Yau condition.} in $X_{N,M}$. In \cite{Hohenegger:2016yuv} it was argued (and reviewed in \cite{Bastian:2017ing}) that a particular sequence of such transformations allows to connect the K\"ahler cones of $X^{(\delta=0)}_{N,M}$ and $X^{(\delta=0)}_{N',M'}$ if $NM=N'M'$ and $\text{gcd}(N,M)=\text{gcd}(N',M')$.\footnote{It was argued in \cite{Bastian:2018dfu}, that these relations imply the existence of an unexpectedly large web of dualities between the six-dimensional gauge theories that are engineered by these web diagrams. More precisely, it was conjectured that theories with gauge group $[U(N)]^M$ are dual to those with $[U(N')]^{M'}$ where $(N,M)$ and $(N',M')$ are related as above.} The K\"ahler cones of $X_{N,M}$ and $X_{N',M'}$ are thus conjectured to be part of a larger, so-called \emph{extended K\"ahler moduli space}. In \cite{Bastian:2017ing}, this conjecture was explicitly verified at the level of the topological string partition function $\mathcal{Z}_{N,M}$ of $X_{N,M}$ for the case of $\gcd(N,M)=1$. The partition function takes the schematic form
\begin{align}
\mathcal{Z}_{N,M}=\sum_{\alpha}\left(\prod_{i=1}^M\prod_{j=1}^N e^{-u_{ij}|\alpha_j^i|}\right)\,\prod_{j=1}^NW_{\alpha_{j+1}^1\ldots \alpha_{j+1}^M}^{\alpha_j^1\ldots\alpha_j^M} (Q_{\text{K\"ahler}})\,,\label{DefPartitionFunction}
\end{align}
where $W_{\alpha_{j+1}^1\ldots \alpha_{j+1}^M}^{\alpha_j^1\ldots\alpha_j^M}$ are a set of universal building blocks, which are labelled by $NM$ integer partitions $\alpha_j^i$ (with $\alpha_{j+N}^i=\alpha_j^i$) and $(u_{ij},Q_{\text{K\"ahler}})$ are a particular parametrisation of the K\"ahler parameters of $X_{N,M}$. For more details (specifically the precise dependence of $W_{\alpha_{j+1}^1\ldots \alpha_{j+1}^M}^{\alpha_j^1\ldots\alpha_j^M} $ on $Q_{\text{K\"ahler}}$), as well as an interpretation of the latter from a gauge theoretical point of view, we refer the reader to \cite{Bastian:2017ing,Bastian:2018dfu} (for the discussion of the topological string partition function of elliptic Calabi-Yau threefolds, see \cite{Huang:2015sta,Klemm:2012sx}). We remark, however, that $\mathcal{Z}_{N,M}$ transforms covariantly with respect to two copies of $SL(2,\mathbb{Z})$, one of which is made manifest in the formulation (\ref{DefPartitionFunction}). The modular parameter of the latter shall be denoted $\rho$ in the following. 

The toric manifolds $X_{N,M}$ engineer six-dimensional gauge theories compactified on a circle $S^1$ with radius $R$, which (in suitable units) is related to $\rho$ as follows:
\begin{align}
&\text{Im}(\rho)=\frac{1}{R}\,.
\end{align}
The five-dimensional limit $R\to 0$ thus corresponds to the limit $\rho\to i\infty$ (one of the cusps of the corresponding $SL(2,\mathbb{Z})$), which is well defined from the perspective of (\ref{DefPartitionFunction}) and can in fact be obtained through an expansion in terms of $Q_\rho =e^{2\pi i \rho}$. Geometrically, the elliptic parameter $\rho$ is associated to an elliptic curve which can be realised as a sum of $N$ $\mathbb{P}^1$'s. Sending $\rho\to i\infty$ is achieved by taking the size of one of these $\mathbb{P}^1$'s to infinity. The result is still a toric Calabi-Yau manifold which can thus be used to engineer a five-dimensional gauge theory.

\begin{figure}
\begin{center}
\scalebox{0.53}{\parbox{16cm}{\begin{tikzpicture}[scale = 1.50]
\draw[ultra thick] (0,0) -- (1,1) -- (2,1) -- (3,2) -- (4,2) -- (4.5,2.5);
\draw[red,ultra thick] (-1,0) -- (0,0);
\draw[red,ultra thick] (8,5) -- (9,5);
\draw[ultra thick, dashed] (4.75,2.75) -- (5.25,3.25);
\draw[ultra thick] (5.5,3.5) -- (6,4) -- (7,4) -- (8,5)  ;
\draw[ultra thick] (1,1) -- (1,2);
\draw[ultra thick] (3,2) -- (3,3);
\draw[ultra thick] (6,4) -- (6,5);
\draw[ultra thick] (8,5) -- (8,6);
\draw[ultra thick] (0,0) -- (0,-1);
\draw[ultra thick] (2,1) -- (2,0);
\draw[ultra thick] (4,2) -- (4,1);
\draw[ultra thick] (7,4) -- (7,3);
\node at (1,2.25) {\large $\mathbf{N+1-\delta}$};
\node at (3,3.25) {\large $\mathbf{N+2-\delta}$};
\node at (6,5.25) {\large $\mathbf{N-1-\delta}$};
\node at (8,6.25) {\large $\mathbf{N-\delta}$};
\node at (0,-1.25) {\large $\mathbf{1}$};
\node at (2,-0.25) {\large $\mathbf{2}$};
\node at (4,0.75) {\large $\mathbf{3}$};
\node at (7,2.75) {\large $\mathbf{N}$};
\node at (-1.25,0) {\large {\bf a}};
\node at (9.25,5) {\large {\bf a}};
\end{tikzpicture}}}
\caption{\sl Web diagram for $X_{N,1}^{(\delta)}$. The labelling of the lines at the top of the diagram is understood to be modulo $N$, such that for $\delta=0$ we get back the 'usual' web diagram $X_{N,1}$.}
\label{Fig:TwistExample}
\end{center}
\end{figure}
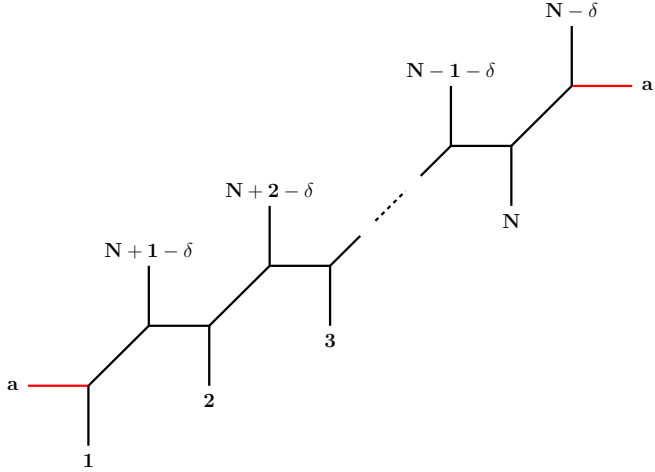

During intermediate steps of the previously mentioned sequence of flop transformations discussed in \cite{Hohenegger:2016yuv}, other K\"ahler cones are encountered which can be described with so-called \emph{shifted} web diagrams. An example of the latter (corresponding to $X_{N,1}^{(\delta)}$ for $\delta\in [0,N-1)$) is shown in \figref{Fig:TwistExample}: the vertical lines at the top and the bottom of the diagram are glued together with a shift $\delta$. In fact, it was argued in \cite{Hohenegger:2016yuv,Bastian:2017ing}, that a $X_{N,M}^{(\delta)}$ web diagram can be related to a $X_{N,M}^{(\delta+k)}$ web diagram (where $k=\gcd(N,M)$) through a combination of flop (and $SL(2,\mathbb{Z})$-) transformations. A five-dimensional limit of the type $\rho\to i\infty$ as in the case of $\delta=0$ before, is difficult to realise for the shifted webs with $\delta\neq 0$. This can be directly understood from the diagram in \figref{Fig:TwistExample}: the na\"ive limit $\rho\to i\infty$ corresponds to pulling the two horizontal lines that are glued along the label $a$ (drawn in red in \figref{Fig:TwistExample}) off to infinity. In this limit (and for $\delta\neq 0$), however, the only way of gluing the lines at the bottom of the diagram to the ones at the top of the diagram involves crossings, which are not consistent. This, however, does not preclude the existence of other, more involved limits of the K\"ahler parameters of the web diagram $X_{N,M}^{(\delta)}$, which leave a consistent web diagram thus engineering a viable five-dimensional gauge theory. In the remainder of this section we shall discuss with the help of a number of examples the existence of such novel limits and analyse the types of five-dimensional gauge theories they can give rise to. While the five-dimensional limit of theories with $\delta=0$ in general leads to the same gauge groups as their six-dimensional counter parts, for $\delta\neq 0$, we find new gauge groups (which are subgroups of their six-dimensional parents), however, with a richer matter content.

\subsection{Example: $(N,M)=(3,1)$}
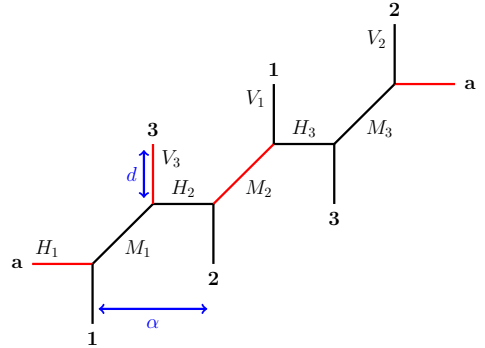
\begin{wrapfigure}{R}{0.4\textwidth}
\begin{center}
\scalebox{0.53}{\parbox{12cm}{\begin{tikzpicture}[scale = 1.50]
\draw[ultra thick] (0,0) -- (1,1)-- (2,1);
\draw[ultra thick] (3,2) -- (4,2) -- (5,3);
\draw[red,ultra thick] (-1,0) -- (0,0);
\draw[red,ultra thick] (5,3) -- (6,3);
\draw[red,ultra thick] (2,1) -- (3,2);
\draw[red,ultra thick] (1,1) -- (1,2);
\draw[ultra thick] (3,2) -- (3,3);
\draw[ultra thick] (5,3) -- (5,4);
\draw[ultra thick] (0,0) -- (0,-1);
\draw[ultra thick] (2,1) -- (2,0);
\draw[ultra thick] (4,2) -- (4,1);
\node at (-0.75,0.25) {\large $H_1$};
\node at (1.5,1.25) {\large $H_2$};
\node at (3.5,2.25) {\large $H_3$};
\node at (1.3,1.75) {\large $V_3$};
\node at (2.7,2.75) {\large $V_1$};
\node at (4.7,3.75) {\large $V_2$};
\node at (0.75,0.25) {\large $M_1$};
\node at (2.75,1.25) {\large $M_2$};
\node at (4.75,2.25) {\large $M_3$};
\node at (1,2.25) {\large $\mathbf{3}$};
\node at (3,3.25) {\large $\mathbf{1}$};
\node at (5,4.25) {\large $\mathbf{2}$};
\node at (0,-1.25) {\large $\mathbf{1}$};
\node at (2,-0.25) {\large $\mathbf{2}$};
\node at (4,0.75) {\large $\mathbf{3}$};
\node at (-1.25,0) {\large {\bf a}};
\node at (6.25,3) {\large {\bf a}};
\draw[ultra thick, <->, blue] (0.1,-0.75) -- (1.9,-0.75);
\node[blue] at (1,-1) {\large $\alpha$};
\draw[ultra thick, <->, blue] (0.85,1.1) -- (0.85,1.9);
\node[blue] at (0.65,1.5) {\large $d$};
\end{tikzpicture}}}
\caption{\sl Web diagram of $X_{3,1}^{(1)}$.}
\label{Fig:3adjTwist}
\end{center}
\end{wrapfigure}
\begin{figure}[h]
\begin{center}
\scalebox{0.53}{\parbox{25.8cm}{\begin{tikzpicture}[scale = 1.50]
\draw[ultra thick] (-1,-1) -- (0,0) -- (2,0) -- (4,-1);
\draw[ultra thick] (0,0) -- (0,1);
\draw[ultra thick] (0,1) -- (1,1) -- (2,0);
\draw[ultra thick] (-3,3) -- (-1,2) -- (0,1);
\draw[ultra thick] (1,1) -- (1,2) -- (2,3);
\draw[ultra thick] (-1,2) -- (1,2);
\node at (-3.2,3.1) {\large  {\bf a}};
\node at (2.2,3.2) {\large  {\bf b}};
\node at (-1.2,-1.2) {\large  {\bf b}};
\node at (4.2,-1.1) {\large  {\bf a}};
\node at (0.35,0.75) {\large  {\bf $B$}};
\node at (0,2.25) {\large  {\bf $F_2+B$}};
\node at (1.35,-0.25) {\large  {\bf $F_1+B$}};
\node at (-0.25,0.5) {\large  {\bf $F_1$}};
\node at (1.7,0.65) {\large  {\bf $F_1$}};
\node at (1.25,1.5) {\large  {\bf $F_2$}};
\node at (-0.7,1.35) {\large  {\bf $F_2$}};
\node at (-0.8,-0.25) {\large  {\bf $-E_1$}};
\node at (1.8,2.3) {\large  {\bf $-E_1$}};
\node at (3.1,-0.25) {\large  {\bf $-E_2$}};
\node at (-2.25,2.3) {\large  {\bf $-E_2$}};
\draw[ultra thick, dashed, red] (0.75,-0.5) -- (0.75,2.5);
\node at (0.5,-3) {\Large  {\bf (a)}};
\begin{scope}[xshift=11cm]
\draw[ultra thick] (-3,5) -- (-1,4) -- (0,3) -- (0,2) -- (-1,1) -- (-1,0) -- (0,-1) -- (2,-2);
\draw[ultra thick] (-1,4) -- (0,4);
\draw[ultra thick] (0,3) -- (1,3);
\draw[ultra thick] (0,2) -- (1,2);
\draw[ultra thick] (-1,1) -- (-2,1);
\draw[ultra thick] (-1,0) -- (-2,0);
\draw[ultra thick] (0,-1) -- (-1,-1);
\node at (-3.2,5.1) {\large  {\bf a}};
\node at (2.2,-2.1) {\large  {\bf a}};
\node at (0.2,4) {\large  {\bf 1}};
\node at (1.2,3) {\large  {\bf 2}};
\node at (1.2,2) {\large  {\bf 3}};
\node at (-2.2,1) {\large  {\bf 1}};
\node at (-2.2,0) {\large  {\bf 2}};
\node at (-1.2,-1) {\large  {\bf 3}};
\node at (-2.25,4.3) {\large  {\bf $-E_2$}};
\node[rotate=35] at (-0.3,4.45) {\large {\bf $F_2+B$}};
\node at (-0.6,3.3) {\large {\bf $F_2$}};
\node at (0.5,3.2) {\large {\bf $B$}};
\node at (-0.3,2.5) {\large {\bf $F_1$}};
\node[rotate=-35] at (0.7,1.5) {\large {\bf $F_1+B$}};
\node at (-0.85,1.7) {\large {\bf $-E_1$}};
\node[rotate=-35] at (-1.7,1.45) {\large {\bf $F_2+B$}};
\node at (-0.75,0.5) {\large {\bf $F_2$}};
\node at (-1.5,0.2) {\large {\bf $B$}};
\node at (-0.3,-0.35) {\large {\bf $F_1$}};
\node[rotate=35] at (-0.7,-1.5) {\large {\bf $F_1+B$}};
\node at (1.1,-1.25) {\large  {\bf $-E_2$}};
\node at (-0.5,-3) {\Large  {\bf (b)}};
\end{scope}
\end{tikzpicture}}}
\caption{\sl (a) Gluing two copies of $\mathbb{F}_1$. (b) Same geometry after cutting along the red line and re-gluing along the line labelled $-E_1$.}
\label{Fig:31StartingF1F1}
\end{center}
\end{figure}
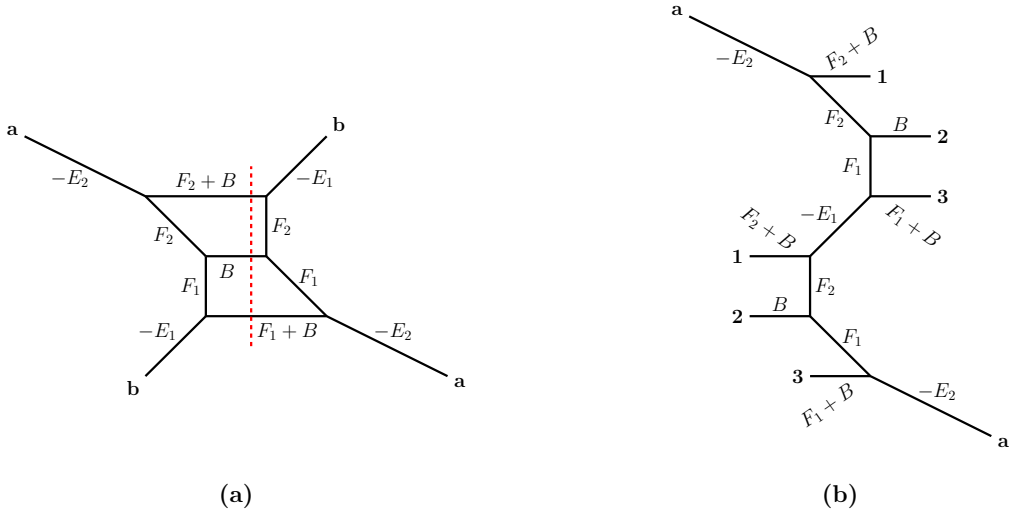
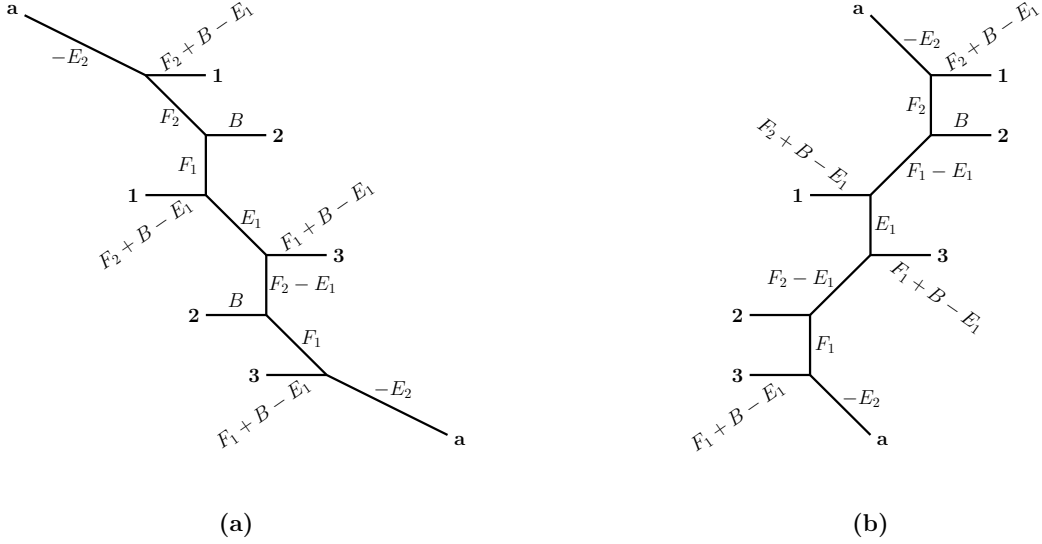
\begin{figure}
\begin{center}
\scalebox{0.53}{\parbox{25.8cm}{\begin{tikzpicture}[scale = 1.50]
\draw[ultra thick] (-3,5) -- (-1,4) -- (0,3) -- (0,2) -- (1,1) -- (1,0) -- (2,-1) -- (4,-2);
\draw[ultra thick] (-1,4) -- (0,4);
\draw[ultra thick] (0,3) -- (1,3);
\draw[ultra thick] (1,1) -- (2,1);
\draw[ultra thick] (-1,2) -- (0,2);
\draw[ultra thick] (0,0) -- (1,0);
\draw[ultra thick] (2,-1) -- (1,-1);
\node at (-3.2,5.1) {\large  {\bf a}};
\node at (4.2,-2.1) {\large  {\bf a}};
\node at (0.2,4) {\large  {\bf 1}};
\node at (1.2,3) {\large  {\bf 2}};
\node at (2.2,1) {\large  {\bf 3}};
\node at (-1.2,2) {\large  {\bf 1}};
\node at (-0.2,0) {\large  {\bf 2}};
\node at (0.8,-1) {\large  {\bf 3}};
\node at (-2.25,4.3) {\large  {\bf $-E_2$}};
\node[rotate=35] at (0,4.65) {\large {\bf $F_2+B-E_1$}};
\node at (-0.6,3.3) {\large {\bf $F_2$}};
\node at (0.5,3.25) {\large {\bf $B$}};
\node at (-0.3,2.5) {\large {\bf $F_1$}};
\node[rotate=35] at (-1,1.35) {\large {\bf $F_2+B-E_1$}};
\node at (0.75,1.6) {\large {\bf $E_1$}};
\node[rotate=35] at (2,1.65) {\large {\bf $F_1+B-E_1$}};
\node at (1.6,0.5) {\large {\bf $F_2-E_1$}};
\node at (0.5,0.25) {\large {\bf $B$}};
\node at (1.75,-0.4) {\large {\bf $F_1$}};
\node[rotate=35] at (0.95,-1.65) {\large {\bf $F_1+B-E_1$}};
\node at (3.1,-1.25) {\large  {\bf $-E_2$}};
\node at (0.5,-3.5) {\Large  {\bf (a)}};
\begin{scope}[xshift=11cm]
\draw[ultra thick] (0,5) -- (1,4) -- (1,3) -- (0,2) -- (0,1) -- (-1,0) -- (-1,-1) -- (0,-2);
\draw[ultra thick] (1,4) -- (2,4);
\draw[ultra thick] (1,3) -- (2,3);
\draw[ultra thick] (0,1) -- (1,1);
\draw[ultra thick] (0,2) -- (-1,2);
\draw[ultra thick] (-1,0) -- (-2,0);
\draw[ultra thick] (-1,-1) -- (-2,-1);
\node at (-0.2,5.1) {\large  {\bf a}};
\node at (0.2,-2.1) {\large  {\bf a}};
\node at (2.2,4) {\large  {\bf 1}};
\node at (2.2,3) {\large  {\bf 2}};
\node at (1.2,1) {\large  {\bf 3}};
\node at (-1.2,2) {\large  {\bf 1}};
\node at (-2.2,0) {\large  {\bf 2}};
\node at (-2.2,-1) {\large  {\bf 3}};
\node at (0.85,4.6) {\large {\bf $-E_2$}};
\node[rotate=35] at (2,4.65) {\large {\bf $F_2+B-E_1$}};
\node at (0.75,3.5) {\large {\bf $F_2$}};
\node at (1.5,3.25) {\large {\bf $B$}};
\node at (1.15,2.4) {\large {\bf $F_1-E_1$}};
\node[rotate=-35] at (-1.1,2.7) {\large {\bf $F_2+B-E_1$}};
\node at (0.25,1.5) {\large {\bf $E_1$}};
\node[rotate=-35] at (1.1,0.3) {\large {\bf $F_1+B-E_1$}};
\node at (-1.15,0.6) {\large {\bf $F_2-E_1$}};
\node at (-0.75,-0.5) {\large {\bf $F_1$}};
\node[rotate=35] at (-2.2,-1.7) {\large {\bf $F_1+B-E_1$}};
\node at (-0.15,-1.4) {\large {\bf $-E_2$}};
\node at (0,-3.5) {\Large  {\bf (b)}};
\end{scope}
\end{tikzpicture}}}
\caption{\sl (a) Geometry of \figref{Fig:31StartingF1F1} after a flop transformation of the line $-E_1$. (b) Same geometry after an $SL(2,\mathbb{Z})$ transformation.}
\label{Fig:31StartingF1F1Flop}
\end{center}
\end{figure}
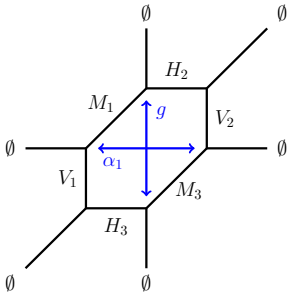
\begin{wrapfigure}{L}{0.3\textwidth}
\begin{center}
\scalebox{0.53}{\parbox{7.3cm}{\begin{tikzpicture}[scale = 1.50]
\draw[ultra thick] (9,0) -- (10,1) -- (11,1) -- (12,2) -- (12,3) -- (13,4);
\draw[ultra thick] (10,1) -- (10,2) -- (11,3) -- (11,4);
\draw[ultra thick] (11,3) -- (12,3);
\draw[ultra thick] (10,2) -- (9,2);
\draw[ultra thick] (11,1) -- (11,0);
\draw[ultra thick] (12,2) -- (13,2);
\draw[ultra thick,<->,blue] (11.8,2) -- (10.2,2);
\draw[ultra thick,<->,blue] (11,2.8) -- (11,1.2);
\node at (11.25,2.6) {\large $\color{blue} g$};
\node at (10.45,1.75) {\large $\color{blue} \alpha_1$};
\node at (10.25,2.75) {\large $M_1$};
\node at (11.7,1.3) {\large $M_3$};
\node at (10.5,0.7) {\large $H_3$};
\node at (11.5,3.3) {\large $H_2$};
\node at (12.3,2.5) {\large $V_2$};
\node at (9.7,1.5) {\large $V_1$};
\node at (8.75,-0.25) {\large $\emptyset$};
\node at (13.25,4.25) {\large $\emptyset$};
\node at (8.75,2) {\large $\emptyset$};
\node at (13.25,2) {\large $\emptyset$};
\node at (11,-0.25) {\large $\emptyset$};
\node at (11,4.25) {\large $\emptyset$};
\end{tikzpicture}}}
\caption{\sl Decompactified web diagram of $X_{3,1}^{(1)}$.}
\label{Fig:3adjTwistDecomp}
\end{center}
\end{wrapfigure}

To illustrate the problem of finding an appropriate decompactification limit for non-trivial shift $\delta\neq 0\text{ mod }N$, we begin with the simplest non-trivial example, namely $X_{3,1}^{(\delta=1)}$, whose web diagram is shown in \figref{Fig:3adjTwist} along with a labelling of all the K\"ahler parameters $(H_{1,2,3},V_{1,2,3},M_{1,2,3})$ (the blue parameters, as well as the significance of the red lines shall be discussed below). Since the vertical curves at the top of the diagram are identified with their counterparts at the bottom with the relative shift $\delta=1$, the limit $H_1\to \infty$ cannot be taken in a way such that none of the remaining lines intersect one another. However, as we shall discuss now, there is another limit which leads to a non-compact web diagram that engineers a five-dimensional gauge theory. To understand how to obtain this limit, we consider the geometry of $\mathbb{F}_1\cup \mathbb{F}_1$ compactified on a torus (\emph{i.e.} with the external legs identified pairwise), as shown in \figref{Fig:31StartingF1F1} (see \cite{Aharony:1997bh} for the non-compact case). The latter can be related to the web diagram of $X_{3,1}^{(\delta=1)}$ in \figref{Fig:3adjTwist} upon performing a flop transformation of the curves $-E_1$ and $-E_2$. To see this, we consider the presentation of the geometry in \figref{Fig:31StartingF1F1} (b) and perform a flop transformation on the curve $-E_1$ to obtain the configuration shown in \figref{Fig:31StartingF1F1Flop} (a). 
After an $SL(2,\mathbb{Z})$ transformation the web diagram takes the form shown in \figref{Fig:31StartingF1F1Flop} (b).
Finally, performing a flop transformation of the curve $-E_2$ we obtain the geometry shown in \figref{Fig:3adjTwist} corresponding to the web diagram of $X_{3,1}^{(\delta=1)}$. Here the K\"ahler parameters are identified
\begin{align}
&H_1=F_2-E_2\,,&&H_2=E_1\,,&&H_3=F_1-E_2\,,\nonumber\\
&V_1=B\,,&&V_2=F_1+B-E_1-E_2\,,&&V_3=F_2+B-E_1-E_2\,,\nonumber\\
&M_1=F_1-E_1\,,&&M_2=F_2-E_1\,,&&M_3=E_2\,.
\end{align}
From the relation between \figref{Fig:31StartingF1F1} (a) and \figref{Fig:3adjTwist} we can get an inspiration for a five-dimensional limit of $X_{3,1}^{(1)}$. Indeed, \figref{Fig:31StartingF1F1} (a) allows various geometrically non-trivial limits, some of which carry over to decompactifications of $X_{3,1}^{(\delta=1)}$ that do not have the problem of crossing or intersecting lines. One example is the limit $F_2\to \infty$\footnote{The latter corresponds to the large fiber limit of one of the $\mathbb{F}_1$ in \figref{Fig:31StartingF1F1} (a).}, which in \figref{Fig:3adjTwist} corresponds to $H_1,V_3,M_2\to \infty$, whose web diagram is shown in \figref{Fig:3adjTwistDecomp}. Here the label $\emptyset$ on the external legs indicates that these lines extend to infinity. To study the five-dimensional  theory associated with this web in

\noindent
more detail (in particular its matter content and partition function) in detail, it is convenient to introduce a basis of parameters $(d,a,g,m_1,m_2)$ (partially shown in \figref{Fig:3adjTwist}), such that all consistency conditions are satisfied
\begin{align}
&2a=\alpha_1=H_2+M_1 \,, &&  g=V_1+M_1\,, && d=V_3 \,, \nonumber\\
&2m_1=M_1-H_2 \,, && 2m_2=M_3-H_3\,.
\end{align}
The relation to the K\"ahler parameters appearing in \figref{Fig:31StartingF1F1} is
\begin{align}
&F_1=2a\,,&&F_2=d+3a-g+m_2\,,&&B=g-a-m_1\,,\nonumber\\
&E_1=a-m_1\,,&&E_2=a+m_2\,.
\end{align}
while in terms of the parameters in \figref{Fig:3adjTwist} we have
\begin{align}
&H_1=2a+d-g\,,&&H_2=a-m_1\,,&&H_3=a-m_2\,,&&M_1=a+m_1\,,\nonumber\\
&M_2=2a+d-g+m_1+m_2\,,&&M_3=a+m_2\,,&&V_1=g-m_1-a\,,&&V_2=g-a-m_2\,,&&V_3=d\,.\nonumber
\end{align}
Using this parametrisation, the partition function $\mathcal{Z}_{3,1}^{(\delta=1)}$ takes the following form: 
\begin{align}\label{pf}
\mathcal{Z}_{3,1}^{(\delta=1)}=&\sum_{\alpha_1,\alpha_2,\alpha_3}    (Q_d Q_aQ_{m_1}^{1/2}Q_{m_2}^{3/2})^{|\alpha_1|}(Q_d Q_a^3Q_{m_1}^{1/2}Q_{m_2}^{3/2})^{|\alpha_2|}(Q_a^{-2}Q_{m_1}Q_{m_2}Q_g)^{|\alpha_3|}\nonumber \\ 
&\times\frac{\ct{1}{1}(\mathcal{Q};\rho)\, \ct{2}{2}(\mathcal{Q};\rho)\, \ct{3}{3}(\mathcal{Q};\rho)\, \ct{1}{2}(\mathcal{Q}\,Q_a^{-2};\rho)\,\ct{2}{1}(\mathcal{Q}\,Q_a^{2};\rho)}{\ct{3}{3}(1;\rho)\, \ct{2}{3}(\mathcal{Q}^{-1}Q_aQ_{m_2};\rho)\, \ct{1}{3}(\mathcal{Q}^{-1}Q_a^{-1}Q_{m_2};\rho)\, \ct{3}{2}(\mathcal{Q}Q_a^{-1}Q_{m_2}^{-1};\rho) } \nonumber \\
&\times \frac{ \ct{2}{3}(Q_aQ_{m_2};\rho) \ct{1}{3}(Q_{m_2}Q_a^{-1};\rho) \ct{3}{2}( Q_{m_1}Q_a^{-1};\rho) \ct{3}{1}(Q_{m_1}Q_a;\rho)}{\ct{3}{1}(\mathcal{Q}Q_aQ_{m_2}^{-1};\rho)\, \ct{1}{1}(1;\rho)\, \ct{2}{2}(1;\rho)\, \ct{1}{2}(Q_a^{-2};\rho)\, \ct{2}{1}(Q_a^{2};\rho)}.
\end{align}
where the modular parameter is  $\rho=8a+2d-2g+m_1+m_2$ and we used the notation 
\begin{align}
&Q_a=e^{-a}\,,&&Q_d=e^{-d}\,,&&Q_g=e^{-g}\,,&&Q_{m_{1,2}}=e^{-m_{1,2}}\,,&&\mathcal{Q}=Q_dQ_a^4Q_{m_1}Q_{m_2}Q_g^{-1}\,,\nonumber
\end{align}
and the summations $\alpha_1\,,\alpha_2$ and $\alpha_3$ are over integer partitions of $|\alpha_1|\,,|\alpha_2|$ and $|\alpha_3|$ respectively. For the notation of the functions $\vartheta_{\mu\nu}$, we refer the reader to \emph{e.g.} \cite{Bastian:2017ing}.

\noindent
In the limit $d\to \infty$ one has
\begin{align}
&H_1\to \infty\,,&&V_3\to \infty\,,&&M_2\to\infty\,,
\end{align}
which implies that the areas of the curves drawn in red in \figref{Fig:3adjTwist} become infinite and the corresponding lines are pulled to infinity. The remaining web diagram (shown in \figref{Fig:3adjTwistDecomp}) is still toric and in fact corresponds to local $dP_3$. The latter engineers \cite{Aharony:1997bh} a five-dimensional gauge theory with gauge group\footnote{ There are no extra contributions to the partition function coming from parallel external legs \cite{Taki:2014pba}. Thus, the gauge group obtained is $SU(2)$ instead of $U(2)$.} $SU(2)$ and whose matter content transforms as $N_f=2$ copies of the fundamental representation. We denote this theory as $(SU(2),2\,\mathbf{F})$.

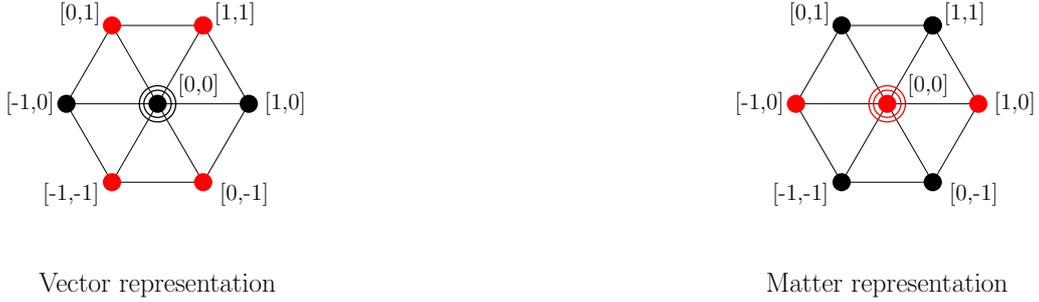
\begin{figure}
\begin{center}
\scalebox{0.6}{\parbox{22.8cm}{\begin{tikzpicture}[scale = 2]
\draw(-1,0) -- (1,0);
\draw (0.5,0.866) -- (-0.5,-0.866);
\draw (-0.5,0.866) -- (0.5,-0.866);
\draw (1,0) -- (0.5,-0.866) -- (-0.5,-0.866) -- (-1,0) -- (-0.5,0.866) -- (0.5,0.866) -- (1,0);

\fill (0,0) circle (0.1cm);
\draw[thick] (0,0) circle (0.15cm);
\draw[thick] (0,0) circle (0.2cm);
\fill (1,0) circle (0.1cm);
\fill (-1,0) circle (0.1cm);
\fill[red] (0.5,0.866) circle (0.1cm);
\fill[red] (-0.5,0.866) circle (0.1cm);
\fill[red] (0.5,0.866) circle (0.1cm);
\fill[red] (-0.5,-0.866) circle (0.1cm);
\fill[red] (0.5,-0.866) circle (0.1cm);
\node at (0,-2) {\Large  {Vector representation}};

\node at (0.85,1) {{\large [1,1]}};
\node at (1.4,0) {{\large [1,0]}};
\node at (-0.85,1) {{\large [0,1]}};
\node at (-1.4,0) {{\large [-1,0]}};
\node at (-0.95,-1) {{\large [-1,-1]}};
\node at (0.95,-1) {{\large [0,-1]}};
\node at (0.45,0.2) {{\large [0,0]}};
\begin{scope}[xshift=8cm]
\draw (-1,0) -- (1,0);
\draw (0.5,0.866) -- (-0.5,-0.866);
\draw (-0.5,0.866) -- (0.5,-0.866);
\draw (1,0) -- (0.5,-0.866) -- (-0.5,-0.866) -- (-1,0) -- (-0.5,0.866) -- (0.5,0.866) -- (1,0);
\fill[red] (0,0) circle (0.1cm);
\draw[thick, red] (0,0) circle (0.15cm);
\draw[thick, red] (0,0) circle (0.2cm);
\fill[red] (1,0) circle (0.1cm);
\fill[red] (-1,0) circle (0.1cm);
\fill (0.5,0.866) circle (0.1cm);
\fill (-0.5,0.866) circle (0.1cm);
\fill (0.5,0.866) circle (0.1cm);
\fill (-0.5,-0.866) circle (0.1cm);
\fill (0.5,-0.866) circle (0.1cm);
\node at (0.85,1) {{\large [1,1]}};
\node at (1.4,0) {{\large [1,0]}};
\node at (-0.85,1) {{\large [0,1]}};
\node at (-1.4,0) {{\large [-1,0]}};
\node at (-0.95,-1) {{\large [-1,-1]}};
\node at (0.95,-1) {{\large [0,-1]}};
\node at (0.45,0.2) {{\large [0,0]}};
\node at (0,-2) {\Large  {Matter representation}};
\end{scope}
\end{tikzpicture}}}
\caption{\sl The vector and matter representation of the $(3,1)$ web with shift $\delta=1$. The weights in red get "projected" out upon taking the $5d$ limit. The circles around the weight $[0,0]$ indicate that the latter is threefold degenerate.}
\label{VecMatterRep}
\end{center}
\end{figure}

In the limit $d\to \infty$, also the partition function (\ref{pf}) simplifies, in particular the theta functions can be written in terms of the Nekrasov factors\footnote{For a partition $\mu=(\mu_{1},\mu_{2},\cdots)$ with parts $\mu_{i}$ we define $||\mu||^2=\sum_{i}\mu_{i}^2$} as follows
\begin{align}
&\vartheta_{\mu\nu}(x;\rho)\,\,\xrightarrow[\rho \to i\infty]\,\, x^{-\frac{|\mu|+|\nu|}{2}}\,t^{\frac{1}{4}(||\mu^t||^2-||\nu^t||^2)}\,q^{-\frac{1}{4}(||\mu||^2-||\nu||^2)}\,N_{\mu\nu}\Big(x\sqrt{\tfrac{t}{q}}\Big)\,,\nonumber\\
&\vartheta_{\mu\nu}(x;\rho)\,\,\xrightarrow[\rho \to i\infty\,,x\mapsto 0]\,\,x^{-\frac{|\mu|+|\nu|}{2}}\,t^{\frac{1}{4}(||\mu^t||^2-||\nu^t||^2)}\,q^{-\frac{1}{4}(||\mu||^2-||\nu||^2)}\,,\label{limit}
\end{align}
where the Nekrasov factor is defined as
\bea
N_{\mu\nu}(x)=\prod_{(i,j)\in \mu}\Big(1-x\,t^{\nu^{t}_{j}-i}\,q^{\mu_{i}-j+1}\Big)\prod_{(i,j)\in \nu}\Big(1-x\,t^{-\mu^{t}_{j}+i-1}\,q^{-\nu_{i}+j}\Big)\,.
\eea
Using (\ref{limit}) in (\ref{pf}) the partition function takes the following form: 
\begin{align}
\mathcal{Z}_{dP_3}^{5D}=&\sum_{\alpha_{1,2}}(-1)^{|\alpha_{1}|+|\alpha_{2}|}\,\Big(Q_{a}^{-2}\,Q_{g}\,Q_{m_{1}}^{-1}\,Q_{m_{2}}^{-1}\Big)^{|\alpha_{1}|+|\alpha_2|}\nonumber\\
&\times \frac{N_{\alpha_{1}\emptyset}(Q_{m_2}Q_{a}^{-1}\sqrt{\tfrac{t}{q}})N_{\emptyset\alpha_{1}}(Q_{m_1}Q_{a}\sqrt{\tfrac{t}{q}})N_{\alpha_{2}\emptyset}(Q_{m_2}Q_{a}\sqrt{\tfrac{t}{q}})N_{\emptyset\alpha_{2}}(Q_{m_1}Q_{a}^{-1}\sqrt{\tfrac{t}{q}})}{N_{\alpha_{1}\alpha_{1}}(\sqrt{\tfrac{t}{q}})N_{\alpha_{2}\alpha_{2}}(\sqrt{\tfrac{t}{q}})N_{\alpha_{1}\alpha_{2}}(Q_{a}^{-2}\sqrt{\tfrac{t}{q}})
N_{\alpha_{2}\alpha_{1}}(Q_{a}^2\sqrt{\tfrac{t}{q}})}\,.
\end{align}
which is indeed the partition function of the five-dimensional ${\cal N}=1$ $SU(2)$ gauge theory with $N_{f}=2$ \cite{Nekrasov:2002qd, Nakajima:2005fg, Awata:2010yy,Taki:2014pba}.
By associating weights to the $\vartheta$-functions in \ref{pf} one can see how the matter and vector representations get reduced upon taking the non-trivial five-dimensional limit discussed above. The weights are assigned based on the intersection numbers of the curves that make up the arguments with the compact divisors $S_i$ \cite{Eso}. In appendix \ref{AppB} we review a method to calculate the intersection numbers. Starting with the vector multiplet contribution (which gives rise to the $\vartheta$-functions in the denominator of the partition function in eq.~(\ref{pf})) we have the following weight assignments in terms of Dynkin labels $([\lambda_1],[\lambda_2])$ of $\mathfrak{su}(3)$:
\begin{align}
&\vartheta_{\alpha_1\alpha_1} \to [0,0] \,, \quad &&\vartheta_{\alpha_2\alpha_2} \to [0,0] \,, \quad &&\vartheta_{\alpha_3\alpha_3} \to [0,0]\nonumber \\
&\vartheta_{\alpha_1\alpha_2} \to [2,-1] \,, \quad
&&\vartheta_{\alpha_2\alpha_3} \to [-1,2] \,, \quad &&\vartheta_{\alpha_1\alpha_3} \to [1,1] \nonumber \\
&\vartheta_{\alpha_2\alpha_1} \to [-2,1] \,, \quad &&\vartheta_{\alpha_3\alpha_2} \to [-1,2] \,, \quad
&&\vartheta_{\alpha_3\alpha_1} \to [-1,-1] 
\end{align}
As expected for a vector multiplet, we find the $\mathfrak{su}(3)$ adjoint representation plus a singlet. We can perform a similar analysis for the adjoint hypermultiplet contribution (which gives rise to the $\vartheta$-functions in the numerator of the partition function in eq.~(\ref{pf})):
\begin{align}
&\vartheta_{\alpha_1\alpha_1} \to [0,0] \,, \quad &&\vartheta_{\alpha_2\alpha_2} \to [0,0] \,, \quad
&&\vartheta_{\alpha_3\alpha_3} \to [0,0] \nonumber \\ &\vartheta_{\alpha_1\alpha_2} \to [2,-1] \,, \quad
&&\vartheta_{\alpha_2\alpha_3} \to [-1,2] \,, \quad &&\vartheta_{\alpha_1\alpha_3} \to [1,1] \nonumber \\
&\vartheta_{\alpha_2\alpha_1} \to [-2,1] \,, \quad &&\vartheta_{\alpha_3\alpha_2} \to [-1,2] \,, \quad
&&\vartheta_{\alpha_3\alpha_1} \to [-1,-1] 
\end{align}
Again, as expected this gives the adjoint representation for the hypermultiplet plus a singlet. We can represent the vector and hypermultiplet representation in the weight lattices, as shown in \figref{VecMatterRep}. The weights that are colored in red, are related to the $\vartheta$-functions whose argument goes to zero upon taking the $5d$ limit. These weights get projected out and we are left with an adjoint $\mathfrak{su}(2)$ representation plus a singlet for the vector multiplet and two fundamental $\mathfrak{su}(2)$ representations for the hypermultiplet. This agrees, as it should, with what one expects at the level of the web-diagram.
\subsection{Example: $(N,M)=(5,1)$}
Similar decompactification limits as for the web diagram of $X_{3,1}^{(\delta=1)}$ discussed in the previous section, can also be found for  other configurations. As the next non-trivial example we present the case $X_{5,1}^{(\delta)}$ for shifts $\delta=2$ and $\delta=3$.\footnote{The remaining shifts $\delta=1$ and $\delta=4$ can be directly related to these two by means of a simple re-arrangement of the web diagram. They therefore do not give rise to new non-trivial limits.}
\subsubsection{Shift $\delta=2$}
\begin{figure}
\begin{center}
\scalebox{0.53}{\parbox{17.75cm}{\begin{tikzpicture}[scale = 1.50]
\draw[ultra thick] (0,0) -- (1,1) -- (2,1) -- (3,2) -- (4,2);
\draw[ultra thick] (5,3) -- (6,3) -- (7,4) -- (8,4) -- (9,5);
\draw[ultra thick,red] (-1,0) -- (0,0);
\draw[ultra thick,red] (9,5) -- (10,5);
\draw[ultra thick, red] (4,2) -- (5,3);
\draw[ultra thick,red] (1,1) -- (1,2);
\draw[ultra thick,red] (3,2) -- (3,3);
\draw[ultra thick] (5,3) -- (5,4);
\draw[ultra thick] (7,4) -- (7,5);
\draw[ultra thick] (9,5) -- (9,6);
\draw[ultra thick] (0,0) -- (0,-1);
\draw[ultra thick] (2,1) -- (2,0);
\draw[ultra thick] (4,2) -- (4,1);
\draw[ultra thick,red] (6,3) -- (6,2);
\draw[ultra thick,red] (8,4) -- (8,3);
\node at (-0.75,0.25) {\large $H_1$};
\node at (1.5,1.25) {\large $H_2$};
\node at (3.5,2.25) {\large $H_3$};
\node at (5.5,3.25) {\large $H_4$};
\node at (7.5,4.25) {\large $H_5$};
\node at (0.75,1.75) {\large $V_4$};
\node at (2.75,2.75) {\large $V_5$};
\node at (5.25,3.75) {\large $V_1$};
\node at (6.75,4.75) {\large $V_2$};
\node at (8.75,5.75) {\large $V_3$};
\node at (0.75,0.25) {\large $M_1$};
\node at (2.75,1.25) {\large $M_2$};
\node at (4.75,2.25) {\large $M_3$};
\node at (6.75,3.25) {\large $M_4$};
\node at (8.75,4.25) {\large $M_5$};
\node at (1,2.25) {\large $\mathbf{4}$};
\node at (3,3.25) {\large $\mathbf{5}$};
\node at (5,4.25) {\large $\mathbf{1}$};
\node at (7,5.25) {\large $\mathbf{2}$};
\node at (9,6.25) {\large $\mathbf{3}$};
\node at (0,-1.25) {\large $\mathbf{1}$};
\node at (2,-0.25) {\large $\mathbf{2}$};
\node at (4,0.75) {\large $\mathbf{3}$};
\node at (6,1.75) {\large $\mathbf{4}$};
\node at (8,2.75) {\large $\mathbf{5}$};
\node at (-1.25,0) {\large {\bf a}};
\node at (10.25,5) {\large {\bf a}};
\draw[ultra thick, <->, blue] (0.1,-0.75) -- (1.9,-0.75);
\node[blue] at (1,-1) {\large $\alpha_1$};
\draw[ultra thick, <->, blue] (2.1,0.25) -- (3.9,0.25);
\node[blue] at (3,0) {\large $\alpha_2$};
\draw[ultra thick, <->, blue] (4.75,3.1) -- (4.75,3.9);
\node[blue] at (4.5,3.5) {\large $g$};
\draw[ultra thick, <->, blue] (9.1,5.25) -- (9.9,5.25);
\node[blue] at (9.5,5.5) {\large $d$};
\end{tikzpicture}}}
\caption{\sl Web diagram of $X_{5,1}^{(\delta=2)}$. The blue parameters provide a (partial) basis for labelling the areas for all curves of the diagram. The curves displayed in red, are decompactified in the limit (\ref{51Limitd2}), \emph{i.e.} their area is sent to infinity.}
\label{Fig:5adjTwist}
\end{center}
\end{figure}
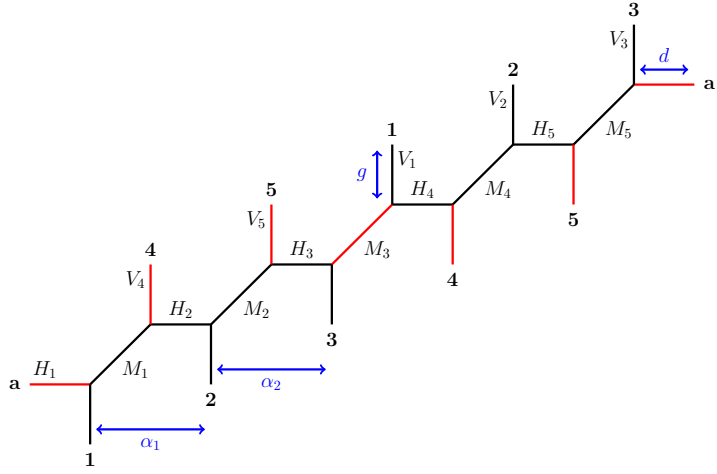

The web diagram of $X_{5,1}^{(2)}$ is shown in \figref{Fig:5adjTwist} along with a suitable labelling of the areas of all individual curves. 
Similar to the previous example, not all of the latter are independent, but they are subject to a number of consistency conditions, as explained in \cite{Haghighat:2013tka,Hohenegger:2015btj}. Inspired by the parametrisation of $X_{3,1}^{(1)}$ above, we can introduce $(d,\alpha_1,\alpha_2,m_1,m_2,m_3,g)$ (as partially indicated in \figref{Fig:5adjTwist}) 
\begin{align}
&d= H_1\,,&&\alpha_1=M_1+H_2 \,, &&\alpha_2=M_2+H_3 \,, && m_1=M_1-H_2\,, \nonumber \\
&m_2=M_2-H_2+M_4-H_5\,, &&m_3= M_5-H_5 \,, &&g=V_2 \,,
\end{align}
which provide a solution to the consistency conditions of the form
{\allowdisplaybreaks
\begin{align}
&H_1=d\,,\hspace{0.5cm}H_2=\frac{\alpha_1-m_1}{2}\,,\hspace{0.5cm}H_3=\frac{\alpha_2-m_2+m_3}{2}\,,\hspace{0.5cm}H_4=\frac{\alpha_1-m_2+m_1}{2}\,,\hspace{0.5cm}H_5=\frac{\alpha_2-m_3}{2}\,,\nonumber\\
&V_1=\frac{2g-2m_1+m_2}{2}\,,\hspace{0.25cm}V_2=g\,,\hspace{0.25cm}V_3=\frac{2g+m_2-2m_3}{2}\,,\hspace{0.25cm}V_4=\frac{2d+2g-m_1+2m_2-2m_3-\alpha_1}{2}\,,\nonumber\\
&V_5=\frac{2d+2g-2m_1+2m_2-m_3-\alpha_2}{2}\,,\hspace{0.75cm}M_1=\frac{m_1+\alpha_1}{2}\,,\hspace{0.75cm}M_2=\frac{m_2-m_3+\alpha_2}{2}\,,\nonumber\\
&M_3=\frac{2d+m_2}{2}\,,\hspace{0.5cm}M_4=\frac{\alpha_1-m_1+m_2}{2}\,,\hspace{0.5cm}M_5=\frac{\alpha_2+m_3}{2}\,.
\end{align}}
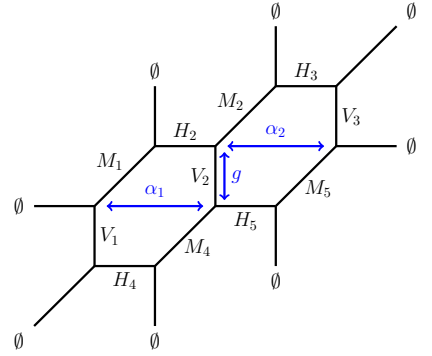
\begin{wrapfigure}{R}{0.33\textwidth}
\begin{center}
\scalebox{0.53}{\parbox{10.4cm}{\begin{tikzpicture}[scale = 1.50]
\draw[ultra thick] (9,0) -- (10,1) -- (11,1) -- (12,2) -- (12,3) -- (13,4) -- (14,4) -- (15,5);
\draw[ultra thick] (10,1) -- (10,2) -- (11,3) -- (11,4);
\draw[ultra thick] (11,3) -- (12,3);
\draw[ultra thick] (10,2) -- (9,2);
\draw[ultra thick] (11,1) -- (11,0);
\draw[ultra thick] (12,2) -- (13,2);
\draw[ultra thick] (13,2) -- (13,1);
\draw[ultra thick] (13,2) -- (14,3) -- (14,4);
\draw[ultra thick] (14,3) -- (15,3);
\draw[ultra thick] (13,4) -- (13,5);
\draw[ultra thick,<->,blue] (11.8,2) -- (10.2,2);
\draw[ultra thick, <->,blue] (13.8,3) -- (12.2,3);
\draw[ultra thick, <->,blue] (12.15,2.1) -- (12.15,2.9);
\node at (11,2.25) {\large $ \color{blue} \alpha_1$};
\node at (13,3.25) {\large $ \color{blue} \alpha_2$};
\node at (12.35,2.5) {\large $ \color{blue} g$};
\node at (10.25,2.75) {\large $M_1$};
\node at (11.5,3.25) {\large $H_2$};
\node at (13.5,4.25) {\large $H_3$};
\node at (12.25,3.75) {\large $M_2$};
\node at (10.5,0.75) {\large $H_4$};
\node at (11.7,1.3) {\large $M_4$};
\node at (12.5,1.75) {\large $H_5$};
\node at (13.7,2.3) {\large $M_5$};
\node at (10.25,1.5) {\large $V_1$};
\node at (11.75,2.5) {\large $V_2$};
\node at (14.25,3.5) {\large $V_3$};
\node at (8.75,-0.25) {\large $\emptyset$};
\node at (15.25,5.25) {\large $\emptyset$};
\node at (8.75,2) {\large $\emptyset$};
\node at (15.25,3) {\large $\emptyset$};
\node at (11,-0.25) {\large $\emptyset$};
\node at (13,5.25) {\large $\emptyset$};
\node at (13,0.75) {\large $\emptyset$};
\node at (11,4.25) {\large $\emptyset$};
\end{tikzpicture}}}
\caption{\sl Web diagram of the decompactified $X_{5,1}^{(\delta=2)}$.}
\label{Fig:SU(2)SU(2)}
\end{center}
${}$\\[-1.5cm]
\end{wrapfigure}

\noindent
In the limit $d \to \infty$ (while keeping the remaining parameters $\alpha_1,\alpha_2,m_1,m_2,m_3,g$ finite) we have 
\begin{align}
&H_1\to \infty\,,&&V_{4,5}\to\infty\,,&& M_3\to\infty\,.\label{51Limitd2}
\end{align}
The corresponding curves in the web diagram \figref{Fig:5adjTwist} (drawn in red) become non-compact and the remaining web diagram can be presented in the form of \figref{Fig:SU(2)SU(2)}. This geometry engineers a five-dimensional gauge theory, which has been previously studied in the literature \cite{Aharony:1997bh}: it corresponds to a theory with gauge group $U(3)$ whose matter content transforms as four copies of its fundamental representation. Therefore, the limit $d\to \infty$ can be interpreted as a decompactification limit of the six-dimensional theory constructed from $X_{5,1}^{(\delta=2)}$ to five dimensions engineering the theory $(U(3),4\,\mathbf{F})$. However, it should be noted that the latter is not at a generic point in the moduli space: the theory $(U(3),4\,\mathbf{F})$ allows for seven independent parameters\footnote{These seven parameters correspond to a gauge coupling constant, two vacuum expectation values of vector multiplet scalars and the masses of the four fundamental matter representations.}, while the theory engineered from the web diagram \figref{Fig:SU(2)SU(2)} only has six independent parameters (as a consequence of the consistency conditions of the compact diagram \figref{Fig:5adjTwist}). As a consequence, the theory engineered in this manner has an additional constraint relating the masses of the matter fields.
\subsubsection{Shift $\delta=3$}
A limit similar to (\ref{51Limitd2}) can also be found for $X_{5,1}^{(\delta=3)}$, whose web diagram (along with a suitable labelling of the areas of the various curves) is shown in \figref{Fig:5adjTwist3}. Inspired by the case $\delta=2$ we introduce the set of parameters $(\alpha_1,\alpha_2,g_1,g_2,m_1,m_2,d)$, which are indirectly defined through
\begin{align}
&\alpha_1=M_1+H_2\,,\hspace{0.8cm}\alpha_2=M_2+H_2\,,\hspace{0.9cm}H_1=d-\alpha_1\,,\nonumber\\
&g_1=V_1+M_1\,,\hspace{1cm}g_2=M_3+V_3\,,\hspace{1cm} m_1=\alpha_2+M_1-H_2\,,\nonumber\\
&m_2=g_1-g_2-H_4+M_3-2\alpha_1+\alpha_2\,.
\end{align}
These parameters provide a solution to the consistency conditions of the web diagram of $X_{5,1}^{(\delta=3)}$ in the sense that the areas of all curves $(H_{1,\ldots,5},V_{1,\ldots,5},M_{1,\ldots,5})$ can be expressed in terms of $(\alpha_1,\alpha_2,g_1,g_2,m_1,m_2,d)$:
\begin{align}
&H_1=d-\alpha_1\,,\hspace{0.75cm}H_2=\frac{\alpha_1+\alpha_2-m_1}{2}\,,\hspace{0.75cm}H_3=d-\alpha_2\,,\hspace{0.75cm}H_4=\alpha_2-\alpha_1+\frac{g_1-g_2-m_2}{2}\,,\nonumber \\
&H_5=\alpha_1-\alpha_2-\frac{g_1-g_2+m_2}{2}\,,\hspace{1cm}V_1=g_1-\frac{m_1+\alpha_1-\alpha_2}{2}\,,\hspace{1cm}V_2=-\alpha_2+\frac{g_1+g_2-m_2}{2}\,,\nonumber\\
&V_3=-\alpha_1+\frac{g_1+g_2-m_2}{2}\,,\hspace{0.25cm}V_4=g_2-\frac{m_1-\alpha_1+\alpha_2}{2}\,,\hspace{0.25cm}V_5=d+\frac{g_1+g_2-m_1+m_2-\alpha_1-\alpha_2}{2}\,,\nonumber\\
&M_1=\frac{m_1+\alpha_1-\alpha_2}{2}\,,\hspace{0.25cm}M_2=\frac{m_1-\alpha_1+\alpha_2}{2}\,,\hspace{0.25cm}M_3=\alpha_1+\frac{m_2-g_1+g_2}{2}\,,\hspace{0.25cm}M_4=d+m_2\,,\nonumber\\
&M_5=\alpha_2+\frac{m_2+g_1-g_2}{2}\,.
\end{align}

\begin{figure}[btp]
\begin{center}
\scalebox{0.53}{\parbox{17.75cm}{\begin{tikzpicture}[scale = 1.50]
\draw[ultra thick] (0,0) -- (1,1) -- (2,1) -- (3,2);
\draw[ultra thick] (4,2) -- (5,3)-- (6,3);
\draw[ultra thick]  (7,4) -- (8,4) -- (9,5);
\draw[ultra thick,red] (-1,0) -- (0,0);
\draw[ultra thick,red] (9,5) -- (10,5);
\draw[ultra thick, red] (3,2) -- (4,2);
\draw[ultra thick, red] (6,3) -- (7,4);
\draw[ultra thick] (1,1) -- (1,2);
\draw[ultra thick] (3,2) -- (3,3);
\draw[ultra thick,red] (5,3) -- (5,4);
\draw[ultra thick] (7,4) -- (7,5);
\draw[ultra thick] (9,5) -- (9,6);
\draw[ultra thick] (0,0) -- (0,-1);
\draw[ultra thick] (2,1) -- (2,0);
\draw[ultra thick] (4,2) -- (4,1);
\draw[ultra thick] (6,3) -- (6,2);
\draw[ultra thick,red] (8,4) -- (8,3);
\node at (-0.75,0.25) {\large $H_1$};
\node at (1.5,1.25) {\large $H_2$};
\node at (3.5,2.25) {\large $H_3$};
\node at (5.5,3.25) {\large $H_4$};
\node at (7.5,4.25) {\large $H_5$};
\node at (0.75,1.75) {\large $V_3$};
\node at (2.75,2.75) {\large $V_4$};
\node at (5.25,3.75) {\large $V_5$};
\node at (6.75,4.75) {\large $V_1$};
\node at (8.75,5.75) {\large $V_2$};
\node at (0.75,0.25) {\large $M_1$};
\node at (2.75,1.25) {\large $M_2$};
\node at (4.75,2.25) {\large $M_3$};
\node at (6.75,3.25) {\large $M_4$};
\node at (8.75,4.25) {\large $M_5$};
\node at (1,2.25) {\large $\mathbf{3}$};
\node at (3,3.25) {\large $\mathbf{4}$};
\node at (5,4.25) {\large $\mathbf{5}$};
\node at (7,5.25) {\large $\mathbf{1}$};
\node at (9,6.25) {\large $\mathbf{2}$};
\node at (0,-1.25) {\large $\mathbf{1}$};
\node at (2,-0.25) {\large $\mathbf{2}$};
\node at (4,0.75) {\large $\mathbf{3}$};
\node at (6,1.75) {\large $\mathbf{4}$};
\node at (8,2.75) {\large $\mathbf{5}$};
\node at (-1.25,0) {\large {\bf a}};
\node at (10.25,5) {\large {\bf a}};
\draw[ultra thick, <->, blue] (0.1,-0.75) -- (1.9,-0.75);
\node[blue] at (1,-1) {\large $\alpha_1$};
\draw[ultra thick, <->, blue] (4.1,1.25) -- (5.9,1.25);
\node[blue] at (5,1) {\large $\alpha_2$};
\end{tikzpicture}}}
\caption{\sl Web diagram of $X_{5,1}^{(\delta=3)}$. The blue parameters provide a (partial) basis for labelling the areas for all curves of the diagram. The curves displayed in red, become non-compact in the limit (\ref{51Limitd3}), \emph{i.e.} their area is sent to infinity.}
\label{Fig:5adjTwist3}
\end{center}
\end{figure}

\begin{wrapfigure}{R}{0.3\textwidth}
\begin{center}
\scalebox{0.53}{\parbox{9cm}{\begin{tikzpicture}[scale = 1.50]
\draw[ultra thick] (9,0) -- (10,1) -- (11,1) -- (12,2) -- (12,3) -- (13,4);
\draw[ultra thick] (10,1) -- (10,2) -- (11,3) -- (11,4) -- (12,5) -- (13,5) -- (14,6);
\draw[ultra thick] (11,3) -- (12,3);
\draw[ultra thick] (10,2) -- (9,2);
\draw[ultra thick] (11,1) -- (11,0);
\draw[ultra thick] (12,2) -- (13,2);
\draw[ultra thick] (13,4) -- (13,5);
\draw[ultra thick] (10,4) -- (11,4);
\draw[ultra thick] (13,4) -- (14,4);
\draw[ultra thick] (12,5) -- (12,6);
\draw[ultra thick,<->,blue] (11.8,2) -- (10.2,2);
\draw[ultra thick, <->,blue] (12.8,4) -- (11.2,4);
\node at (11,2.25) {\large $ \color{blue} \alpha_1$};
\node at (12,4.25) {\large $ \color{blue} \alpha_2$};
\node at (10.25,2.75) {\large $M_1$};
\node at (11.5,3.25) {\large $H_2$};
\node at (12.7,3.3) {\large $M_2$};
\node at (10.5,0.75) {\large $H_5$};
\node at (11.7,1.3) {\large $M_5$};
\node at (9.75,1.5) {\large $V_1$};
\node at (12.25,2.5) {\large $V_2$};
\node at (11.25,4.75) {\large $M_3$};
\node at (12.5,5.25) {\large $H_4$};
\node at (10.75,3.5) {\large $V_3$};
\node at (13.25,4.5) {\large $V_4$};
\node at (8.75,-0.25) {\large $\emptyset$};
\node at (12,6.25) {\large $\emptyset$};
\node at (8.75,2) {\large $\emptyset$};
\node at (14.25,4) {\large $\emptyset$};
\node at (11,-0.25) {\large $\emptyset$};
\node at (14.25,6.25) {\large $\emptyset$};
\node at (13.25,2) {\large $\emptyset$};
\node at (9.75,4) {\large $\emptyset$};
\end{tikzpicture}}}
\caption{\sl Web diagram of the decompactified $X_{5,1}^{(\delta=3)}$.}
\label{Fig:SU(2)SU(2)2}
\end{center}
${}$\\[-1.5cm]
\end{wrapfigure}
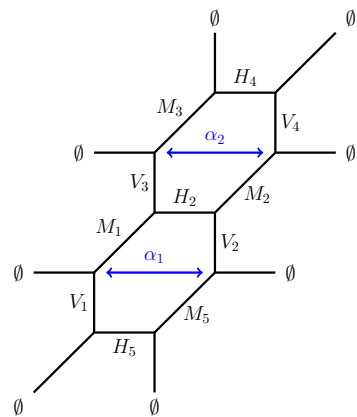

\noindent
In the limit $d\to \infty$ (while keeping the remaining parameters $(\alpha_1,\alpha_2,g_1,g_2,m_1,m_2)$ finite) one finds
\begin{align}
&H_1,H_3\to \infty\,,&&V_5\to\infty\,,&&M_4\to\infty\,,
\label{51Limitd3}
\end{align}
which means that the corresponding curves in the web diagram (indicated in red in \figref{Fig:5adjTwist3}) are decompactified. The resulting diagram can be presented in the form of \figref{Fig:SU(2)SU(2)2}, where $\emptyset$ 

\noindent
indicates that these curves are pulled all the way to infinity. Therefore, the diagram in \figref{Fig:SU(2)SU(2)2} corresponds to a non-compact web and the limit $d\to \infty$ can be understood as a decompactification limit to five dimensions. Furthermore, since  \figref{Fig:SU(2)SU(2)2} can be interpreted as an $SL(2,\mathbb{Z})$ transformation of \figref{Fig:SU(2)SU(2)}, the five-dimensional gauge theory engineered by the former is in fact an S-dual version of $(U(3),4\mathbf{F})$, namely $([U(2)]^2,2\mathbf{F},\mathbf{BF})$. The latter theory is indeed engineered by the same diagram \figref{Fig:SU(2)SU(2)2} up to an $SL(2,\mathbb{Z})$ transformation (which acts essentially as a rotation). 
 
 Finally, similar to the case of  $(U(3),4\mathbf{F})$ engineered from \figref{Fig:SU(2)SU(2)}, the $([U(2)]^2,2\mathbf{F},\mathbf{BF})$ theory engineered from \figref{Fig:SU(2)SU(2)2} is not at a generic point in its moduli space.  Instead, due to the consistency conditions of the original diagram \figref{Fig:5adjTwist3}, the theory is in a co-dimension one subspace. The additional constraint relates the two coupling constants of the gauge groups. 
\section{A Two Parameter Series of Five-dimensional Theories}
\label{TwoPara}
The examples of the previous section can be generalised to more sophisticated configurations: in the following we shall provide non-trivial evidence that five-dimensional gauge theories of the type $\left([U(A+1)]^B,2A\,\mathbf{F}, (B-1)\,\mathbf{BF}\right)$ (with  $A,B\in\mathbb{N}$) can be engineered from a shifted web diagram.

\begin{figure}[h!]
\begin{center}
\scalebox{0.45}{\parbox{39cm}{\begin{tikzpicture}[scale = 1.50]
\node at (-1.25,0) {\large {\bf a}};
\draw[ultra thick] (0,0) -- (1,1) -- (2,1);
\draw[ultra thick,red] (-1,0) -- (0,0);
\draw[ultra thick] (1,1) -- (1,2);
\node at (1,2.25) {\large $\mathbf{A+2}$};
\draw[ultra thick] (0,0) -- (0,-1);
\node at (0,-1.25) {\large $\mathbf{1}$};
\node at (2.5,1) {\Huge $\cdots$};
\draw[ultra thick] (3,1) -- (4,1) -- (5,2); 
\draw[ultra thick,red] (5,2) -- (6,2);
\draw[ultra thick] (6,2) -- (7,3) -- (8,3);
\draw[ultra thick] (5,2) -- (5,3);
\node at (5,3.25) {\large $\mathbf{2(A+1)}$};
\draw[ultra thick] (7,3) -- (7,4);
\node at (7,4.25) {\large $\mathbf{2(A+1)+1}$};
\draw[ultra thick] (4,1) -- (4,0);
\node at (4,-0.25) {\large $\mathbf{A+1}$};
\draw[ultra thick] (6,2) -- (6,1);
\node at (6,0.75) {\large $\mathbf{A+2}$};
\node at (8.5,3) {\Huge $\cdots$};
\draw[ultra thick] (9,3) -- (10,3) -- (11,4) -- (12,4);
\draw[ultra thick,red] (11,4) -- (11,5);
\node at (11,5.25) {\large $\mathbf{B+AB+1}$};
\draw[ultra thick] (10,3) -- (10,2);
\node at (10,1.75) {\large $\mathbf{B+AB-A}$};
\node at (12.5,4) {\Huge $\cdots$};
\draw[ultra thick] (13,4) -- (14,4) -- (15,5) -- (16,5);
\draw[ultra thick,red] (16,5) -- (17,6);
\draw[ultra thick] (17,6) -- (18,6) -- (19,7) -- (20,7);
\draw[ultra thick,red] (15,5) -- (15,6);
\node at (15,6.25) {\large $\mathbf{A+B+AB}$};
\draw[ultra thick] (17,6) -- (17,7);
\node at (17,7.25) {\large $\mathbf{1}$};
\draw[ultra thick] (19,7) -- (19,8);
\node at (19,8.25) {\large $\mathbf{2}$};
\draw[ultra thick] (14,4) -- (14,3);
\node at (14,2.75) {\large $\mathbf{B+AB-1}$};
\draw[ultra thick] (16,5) -- (16,4);
\node at (16,3.75) {\large $\mathbf{B+AB}$};
\draw[ultra thick,red] (18,6) -- (18,5);
\node at (18,4.75) {\large $\mathbf{B+AB+1}$};
\node at (20.5,7) {\Huge $\cdots$};
\draw[ultra thick] (21,7) -- (22,7) -- (23,8);
\draw[ultra thick,red] (23,8) -- (24,8);
\draw[ultra thick] (23,8) -- (23,9);
\node at (23,9.25) {\large $\mathbf{A+1}$};
\draw[ultra thick,red] (22,7) -- (22,6);
\node at (22,5.75) {\large $\mathbf{A+B+AB}$};
\node at (24.25,8) {\large {\bf a}};
\node at (-0.5,0.25) {\large $\mathbf{H_1}$};
\node at (1.5,1.25) {\large $\mathbf{H_2}$};
\node at (3.5,1.25) {\large $\mathbf{H_{A+1}}$};
\node at (5.5,2.25) {\large $\mathbf{H_{A+2}}$};
\node at (7.5,3.25) {\large $\mathbf{H_{A+3}}$};
\node at (9.3,3.25) {\large $\mathbf{H_{B+AB-A}}$};
\node[rotate=45] at (11.85,4.725) {\large $\mathbf{H_{B+AB-A+1}}$};
\node at (13.3,4.25) {\large $\mathbf{H_{B+AB-1}}$};
\node[rotate=45] at (15.75,5.5) {\large $\mathbf{H_{B+AB}}$};
\node[rotate=45] at (17.85,6.6) {\large $\mathbf{H_{B+AB+1}}$};
\node[rotate=45] at (19.85,7.6) {\large $\mathbf{H_{B+AB+2}}$};
\node at (21.3,7.25) {\large $\mathbf{H_{A+B+AB}}$};
\node at (23.5,8.25) {\large $\mathbf{H_{1}}$};
\node at (0.75,0.25) {\large $\mathbf{M_1}$};
\node[rotate=45] at (4.65,1.35) {\large $\mathbf{M_{A+1}}$};
\node[rotate=45] at (6.65,2.35) {\large $\mathbf{M_{A+2}}$};
\node[rotate=45] at (10.65,3.35) {\large $\mathbf{M_{B+AB-A}}$};
\node[rotate=45] at (14.65,4.35) {\large $\mathbf{M_{B+AB-1}}$};
\node[rotate=45] at (16.65,5.35) {\large $\mathbf{M_{B+AB}}$};
\node[rotate=45] at (18.65,6.35) {\large $\mathbf{M_{B+AB+1}}$};
\node[rotate=45] at (22.65,7.35) {\large $\mathbf{M_{A+B+AB}}$};
\node at (0.3,-0.5) {\large $\mathbf{V_{1}}$};
\node at (3.5,0.5) {\large $\mathbf{V_{A+1}}$};
\node at (6.5,1.5) {\large $\mathbf{V_{A+2}}$};
\node at (9.15,2.5) {\large $\mathbf{V_{B+AB-A}}$};
\node at (13.2,3.5) {\large $\mathbf{V_{B+AB-1}}$};
\node at (15.4,4.2) {\large $\mathbf{V_{B+AB}}$};
\node at (18.6,5.4) {\large $\mathbf{V_{B+AB}}$};
\node at (21.1,6.5) {\large $\mathbf{V_{A+B+AB}}$};
\node[blue] at (18,7.3) {\large $S_{1,1}$};
\node[blue] at (20,8.3) {\large $S_{2,1}$};
\node[blue] at (15,3.4) {\large $S_{A,B}$};
\node[blue] at (13,2.4) {\large $S_{A-1,B}$};
\node[blue] at (4,2.4) {\large $S_{A,2}$};
\node[blue] at (1,-0.5) {\large $S_{1,1}$};
\end{tikzpicture}}}
\caption{\sl Web Diagram of $X_{A+B+AB,1}^{(\delta=B(A+1)-1)}$. The curves drawn in red are being decompactified in the limit eq.(\ref{LimitGenWeb}). The blue labels $S_{a,b}$ indicate the same hexagons as in \figref{Fig:ABstack}.}
\label{Fig:NadjTwist}
\end{center}
\end{figure}
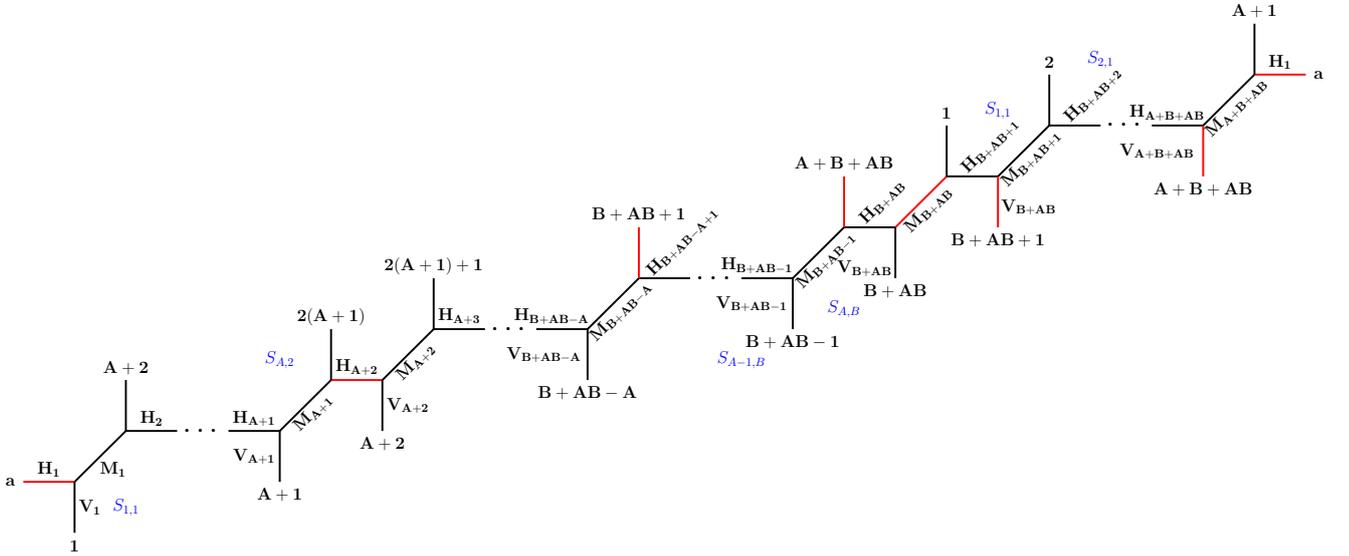

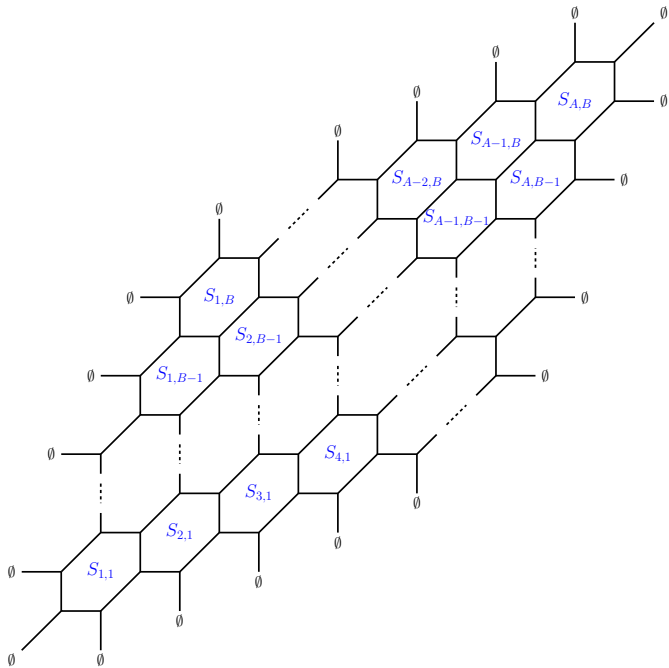
\begin{figure}
\begin{center}
\scalebox{0.4}{\parbox{23.5cm}{\begin{tikzpicture}[scale = 1.30]
\draw[ultra thick] (14.5,1.5) -- (14,1) -- (13,1) -- (12,0) -- (11,0) -- (10,-1) -- (9,-1) -- (9,0) -- (10,1) -- (11,1) -- (12,2) -- (12,3) -- (12.5,3.5);
\draw[ultra thick] (13.5,4.5) -- (14,5) -- (15,5) -- (16,6) -- (17,6) -- (18,7);
\draw[ultra thick] (15,5) -- (15,4) -- (16,4) -- (17,5) -- (18,5);
\draw[ultra thick] (17,5) -- (17,6);
\draw[ultra thick] (8,0) -- (9,0);
\draw[ultra thick] (9,-1) -- (8,-2) -- (8,-2.5);
\draw[ultra thick] (10,-1) -- (10,-1.5);
\draw[ultra thick] (12,0) -- (12,-0.5);
\draw[ultra thick] (14,1) -- (14,0.5);
\draw[ultra thick] (15.5,2.5) -- (16,3) -- (17,3);
\draw[ultra thick] (10,1) -- (10,2) -- (11,3) -- (11,4);
\draw[ultra thick] (11,3) -- (12,3);
\draw[ultra thick] (10,2) -- (9,2);
\draw[ultra thick] (11,1) -- (11,0);
\draw[ultra thick] (12,2) -- (13,2);
\draw[ultra thick] (13,2) -- (13,1);
\draw[ultra thick] (13,2) -- (13.5,2.5);
\draw[ultra thick, dashed] (12.75,3.75) -- (13.25,4.25);
\draw[ultra thick, dashed] (13.75,2.75) -- (14.25,3.25);
\draw[ultra thick, dashed] (14.75,1.75) -- (15.25,2.25);
\draw[ultra thick, dashed] (8,-2.75) -- (8,-3.25);
\draw[ultra thick, dashed] (10,-1.75) -- (10,-2.25);
\draw[ultra thick, dashed] (12,-0.75) -- (12,-1.25);
\draw[ultra thick, dashed] (14,0.25) -- (14,-0.25);
\draw[ultra thick, dashed] (15.75,-0.25) -- (16.25,0.25);
\draw[ultra thick, dashed] (17,2.25) -- (17,1.75);
\draw[ultra thick, dashed] (16.75,-1.25) -- (17.25,-0.75);
\draw[ultra thick, dashed] (19,2.75) -- (19,3.25);
\draw[ultra thick] (14,-0.5) -- (14,-1) -- (13,-2) -- (12,-2) -- (11,-3) -- (10,-3) -- (9,-4) -- (8,-4) -- (7,-5);
\draw[ultra thick] (17,3) -- (18,4) -- (18,5) -- (19,6) -- (20,6);
\draw[ultra thick] (19,3.5) -- (19,4) -- (20,5) -- (20,6) -- (21,7) -- (21,8) -- (22,9);
\draw[ultra thick] (21,8) -- (20,8) -- (20,9);
\draw[ultra thick] (21,7) -- (22,7);
\draw[ultra thick] (20,5) -- (21,5);
\draw[ultra thick] (19,2) -- (20,2);
\draw[ultra thick] (18,0) -- (19,0);
\draw[ultra thick] (19,6) -- (19,7) -- (20,8);
\draw[ultra thick] (19,7) -- (18,7) -- (18,8);
\draw[ultra thick] (18,4) -- (19,4) ;
\draw[ultra thick] (17,1.5) -- (17,1) ;
\draw[ultra thick] (7,-5) -- (7,-6) -- (8,-6) -- (9,-5) -- (10,-5) -- (11,-4) -- (12,-4) -- (13,-3) -- (14,-3) -- (15,-2) -- (16,-2) -- (16.5,-1.5);
\draw[ultra thick] (7,-5) -- (6,-5) ;
\draw[ultra thick] (7,-6) -- (6,-7) ;
\draw[ultra thick] (8,-6) -- (8,-7) ;
\draw[ultra thick] (10,-5) -- (10,-6) ;
\draw[ultra thick] (12,-4) -- (12,-5) ;
\draw[ultra thick] (14,-3) -- (14,-4) ;
\draw[ultra thick] (16,-2) -- (16,-3) ;
\draw[ultra thick] (17,3) -- (17,2.5) ;
\draw[ultra thick] (17.5,-0.5) -- (18,0) -- (18,1) -- (19,2) -- (19,2.5);
\draw[ultra thick] (16.5,0.5) -- (17,1) -- (18,1);
\draw[ultra thick] (8,-4) -- (8,-3.5);
\draw[ultra thick] (10,-3) -- (10,-2.5);
\draw[ultra thick] (12,-2) -- (12,-1.5);
\draw[ultra thick] (9,-4) -- (9,-5);
\draw[ultra thick] (11,-3) -- (11,-4);
\draw[ultra thick] (13,-2) -- (13,-3);
\draw[ultra thick] (15,-1) -- (15,-2);
\draw[ultra thick] (14,-1) -- (15,-1) -- (15.5,-0.5);
\draw[ultra thick] (14.5,3.5) -- (15,4);
\draw[ultra thick] (16,4) -- (16,3);
\draw[ultra thick] (14,5) -- (14,6);
\draw[ultra thick] (16,6) -- (16,7);
\draw[ultra thick] (7,-2) -- (8,-2);
\node at (16,7.25) {\large $\emptyset$};
\node at (18,8.25) {\large $\emptyset$};
\node at (20,9.25) {\large $\emptyset$};
\node at (22.25,9.25) {\large $\emptyset$};
\node at (21.25,5) {\large $\emptyset$};
\node at (20.25,2) {\large $\emptyset$};
\node at (19.25,0) {\large $\emptyset$};
\node at (22.25,7) {\large $\emptyset$};
\node at (8.75,2) {\large $\emptyset$};
\node at (14,6.25) {\large $\emptyset$};
\node at (11,4.25) {\large $\emptyset$};
\node at (16,-3.25) {\large $\emptyset$};
\node at (14,-4.25) {\large $\emptyset$};
\node at (12,-5.25) {\large $\emptyset$};
\node at (10,-6.25) {\large $\emptyset$};
\node at (8,-7.25) {\large $\emptyset$};
\node at (5.75,-7.25) {\large $\emptyset$};
\node at (5.75,-5) {\large $\emptyset$};
\node at (6.75,-2) {\large $\emptyset$};
\node at (7.75,0) {\large $\emptyset$};
\node[blue] at (8,-5) {\Large $S_{1,1}$};
\node[blue] at (10,-4) {\Large $S_{2,1}$};
\node[blue] at (12,-3) {\Large $S_{3,1}$};
\node[blue] at (14,-2) {\Large $S_{4,1}$};
\node[blue] at (10,0) {\Large $S_{1,B-1}$};
\node[blue] at (12,1) {\Large $S_{2,B-1}$};
\node[blue] at (17,4) {\Large $S_{A-1,B-1}$};
\node[blue] at (19,5) {\Large $S_{A,B-1}$};
\node[blue] at (11,2) {\Large $S_{1,B}$};
\node[blue] at (16,5) {\Large $S_{A-2,B}$};
\node[blue] at (18,6) {\Large $S_{A-1,B}$};
\node[blue] at (20,7) {\Large $S_{A,B}$};
\end{tikzpicture}}}
\caption{\sl Web diagram consisting of $AB$ hexagons obtained in the decompactification limit (\ref{LimitGenWeb}) from \figref{Fig:NadjTwist}.}
\label{Fig:ABstack}
\end{center}
\end{figure}

Our starting point is the web diagram of $X_{A+B+AB,1}^{(\delta=B(A+1)-1)}$, as shown in \figref{Fig:NadjTwist}, where we have added a labelling of the areas of all curves. Here we assume $A>0$ and $B>0$. Web diagrams of this type are parametrised by $A+B+AB+2$ independent parameters. Generalising the pattern of the previous examples, we conjecture that there exists a one-parameter limit such that
\begin{align}
&H_{1+k(A+1)}\to \infty\,,&&V_{B+AB+r}\to\infty\,,&&M_{B+AB}\to \infty&&\forall \left\{\begin{array}{l}k=0,\ldots,B-1\\ r=1,\ldots,A\end{array}\right.\label{LimitGenWeb}
\end{align}
while keeping the areas of all remaining curves finite. The $B$ horizontal, $A$ vertical and the single diagonal curve that are decompactified in the limit (\ref{LimitGenWeb}) are shown in red in \figref{Fig:NadjTwist}. The resulting web diagram is non-compact and can be presented in the form of \figref{Fig:ABstack}. The latter corresponds to $AB$ hexagons $S_{a,b}$ (with $a=1,\ldots,A$ and $b=1,\ldots,B$) being arranged in a symmetric array.

The web diagram \figref{Fig:ABstack} engineers a (linear) quiver gauge theory in five dimensions with $B$ gauge nodes of gauge group $U(A+1)$, whose matter content consists of $B-1$ bifundamental representations and $N_f=2A$ fundamental representations, which we denote by $\left([U(A+1)]^B,2A\,\textbf{F}\,,(B-1)\,\textbf{BF}\right)$. The latter can be interpreted as the five-dimensional limit of the six-dimensional theory with gauge group $U(A+B+AB)$ and adjoint matter that is engineered from  $X_{A+B+AB,1}^{(\delta=B(A+1)-1)}$. It should however be noticed that the five-dimensional theory is not at a generic point in its moduli space: this can be seen by counting the independent parameters of the web diagrams. The web diagram in \figref{Fig:NadjTwist} has a maximal set of $A+B+AB+2$ parameters and the decompactification limit gets rid of one of them. As a consequence, we are left with $A+B+AB+1$ independent parameters in \figref{Fig:ABstack}. However, the most generic web of the type \figref{Fig:ABstack} allows for $AB+2(A+B)-1$ parameters. Hence, the theories we obtain in the reduction from six dimensions live in a codimension $A+B-2$ subspace of the full moduli space.
For low values of $N=A+B+AB$, some of the five-dimensional theories that can be engineered as limits of shifted web diagrams are displayed in table \ref{5dTable}.
\begin{table}[ht]
\begin{center}
\begin{tabular}{c||c|c|c||c|c|c}
& \multicolumn{3}{c||}{parameters of $X_{A+B+AB,1}^{(\delta=B(A+1)-1)}$} & \multicolumn{3}{c}{5-dim. gauge theory}  \\[2pt]\hline\hline
 $N$   & $A$ & $B$ & $\delta=B(A+1)-1$ & gauge group & {\bf F} & {\bf BF}\\\hline
 $3$   & $1$ & $1$ & $1$ & $SU(2)$ & $2$ & 0\\\hline
 $5$   & $2$ & $1$ & $2$ & $U(3)$ & $4$ & 0\\
   & $1$ & $2$ & $3$ & $[U(2)]^2$ & $2$ & 1\\\hline
 $7$   & $3$ & $1$ & $3$ & $U(4)$ & $6$ & 0\\
   & $1$ & $3$ & $5$ & $[U(2)]^3$ & $2$ & 2\\\hline
 $8$   & $2$ & $2$ & $5$ & $[U(3)]^2$ & $4$ & 1\\\hline
 $9$   & $4$ & $1$ & $4$ & $U(5)$ & $8$ & 0\\
  & $1$ & $4$ & $7$ & $[U(2)]^4$ & $2$ & 3\\\hline
 $11$   & $5$ & $1$ & $5$ & $U(6)$ & $10$ & 0\\
  & $3$ & $2$ & $7$ & $[U(4)]^2$ & $6$ & 1\\
  & $2$ & $3$ & $8$ & $[U(3)]^3$ & $4$ & 2\\
  & $1$ & $5$ & $9$ & $[U(2)]^5$ & $2$ & 4\\\hline
 $13$   & $6$ & $1$ & $6$ & $U(7)$ & $12$ & 0\\
  & $1$ & $6$ & $11$ & $[U(2)]^6$ & $2$ & 5\\\hline
$14$   & $4$ & $2$ & $9$ & $[U(5)]^2$ & $8$ & 1\\
& $2$ & $4$ & $11$ & $[U(3)]^4$ & $4$ & 3\\\hline
$15$   & $7$ & $1$ & $7$ & $U(8)$ & $14$ & 0\\
   & $3$ & $3$ & $11$ & $[U(4)]^3$ & $6$ & 2\\
  & $1$ & $7$ & $13$ & $[U(2)]^7$ & $2$ & 6\\
\end{tabular}
\end{center}
\caption{Non-trivial five-dimensional limits for $N\leq15$}
\label{5dTable}
\end{table}
\noindent
It should be noted that web diagrams of the same 'length' $N$ (but different shift $\delta$) can give rise to five-dimensional gauge theories of different rank. We can also describe the five dimensional theories that can be obtained from a six-dimensional $U(N)$ theory in terms of the parameter $N$. In this description the possible five-dimensional theories include $[U(\frac{N+1}{D})]^{D-1}$ with $D-2$ bifundamentals and $2(\frac{N+1}{D}-1)$ fundamentals, where $D$ is any positive non-trivial divisor of $N+1$. This presentation explains various gaps in table \ref{5dTable}. Indeed when $N+1$ is a prime number there are no non-trivial divisors and hence no five-dimensional limit of the type discussed above.

\section{Conclusions}\label{Sect:Conclusions} 
In this paper, we continued our exploration of the extended K\"ahler moduli space of a class of toric Calabi-Yau threefolds $X_{N,1}$. In previous work \cite{Hohenegger:2015btj} we have shown that the latter engineers a $U(N)$ gauge theory with adjoint matter on $\mathbb{R}^5\times S^1$, which in the limit of vanishing $S^1$-radius gives rise to a five-dimensional theory with the same gauge group. In this work, we investigated regions in the moduli space of $X_{N,1}$, where the geometry is described by so-called shifted web diagrams $X_{N,1}^{(\delta)}$ (with $\delta\in \{0,\dots,N-1\}$) first introduced in \cite{Hohenegger:2016yuv}, an example of which is shown in \figref{Fig:TwistExample}. We specifically investigated  co-dimension one limits in which the latter engineer five-dimensional quiver gauge theories with gauge group $G\subset U(N)$ (and various different matter contents). At the level of the six-dimensional theories, the topological string partition functions computed from $X_{N,1}^{(\delta)}$ and $X_{N,1}^{(0)}=X_{N,1}$ are identical  (as shown in \cite{Bastian:2017ing}) and thus engineer dual six-dimensional gauge theories, which, however, admit different five-dimensional limits. 

Our analysis is based on a number of specific examples, from which we managed to extract an emergent pattern, which we conjecture to hold in general: starting from a six-dimensional $\mathcal{N}=(1,0)$ $U(N)$ gauge theory with adjoint matter, we conjecture that there exists a five-dimensional limit engineering an $\mathcal{N}=1$ $[U(\frac{N+1}{D})]^{D-1}$ gauge theory with matter transforming in $D-2$ copies of the bifundamental and $2(\frac{N+1}{D}-1)$ copies of the fundamental representation, where $D$ is a positive non-trivial divisor of $N+1$. Furthermore, the limits can already be described at the level of the web diagram, which allows for a direct computation of the partition function using the (refined) topological vertex (as showcased in the example $X_{3,1}^{(1)}$, where the five-dimensional limit gives rise to the web diagram of $dP_3$).

The limits we studied in this paper certainly do not exhaust all possible ways to construct five-dimensional gauge theories from parent theories on $\mathbb{R}^5\times S^1$ that are engineered from (shifted) web diagrams. As an example, several webs allow for limits in which the six-dimensional gauge group $U(N)$ is completely broken to $[U(1)]^N$, which, however, were not discussed in the present work since they are rather trivial. We, however, cannot rule out the existence of further limits, which give rise to five-dimensional limits with a non-trivial gauge group and leave the study of their existence to future work. Conversely, the five-dimensional theories obtained in this work are consistent with the classification of five-dimensional theories discussed in \cite{Jefferson:2018irk} as all of them appear via geometric engineering from Calabi-Yau threefolds. 

\section*{Acknowledgements}
A.I. would like to thank the hospitality of the Simons Center for Geometry and Physics
during the 2018 Summer Workshop in Mathematics and Physics.
\appendix
\section{Decompactifying limit and the partition function}
\label{AppA}
In this section we discuss the relation between taking the limit $Q_{\star}\mapsto0$ (at the level of the partition function $\mathcal{Z}_{N,M}$) and the corresponding effect on the web diagram of $X^{(\delta)}_{N,M}$ (with a potential shift $\delta\neq 0$). We shall keep the discussion completely generic and consider $Q_{\star}=e^{-t_{\star}}$ where $t_{\star}$ is the K\"ahler parameter such that (keeping all other parameters fixed) it controls the size of a certain number of $\mathbb{P}^{1}$'s in the geometry. We would like to show that in the limit $Q_{\star}\mapsto 0$ the topological string partition function remains well defined and is given by a corresponding decompactification of the web diagram. Let $W$ be the web diagram and denote the corresponding partition function by $Z_{W}$. In the limit $Q_{\star}\mapsto 0$ suppose that the web diagram becomes $W_{0}$ with partition function $Z_{W_{0}}$. Recall that the topological string partition function written with the help of the topological vertex is an expansion near the large K\"ahler structure limit \cite{Bershadsky:1993cx} and  are given by a power series in $e^{-t_{a}}$ where $t_{a}$ are the K\"ahler parameters. In the case at hand, we keep the K\"ahler parameters other than $t_{\star}$ fixed as we take the limit $ t_{\star} \mapsto \infty $

We consider the partition function $Z_{W}/Z_{W_0}$ as an expansion in $Q_{\star}$:
\begin{align}
\frac{Z_{W}}{Z_{W_{0}}}\sum_{\nu_{1}\cdots \nu_{k}}Q_{\star}^{|\nu_{1}|+\cdots +|\nu_{k}|}\,\frac{Z_{\nu_{1}\cdots \nu_{k}}(Q_{i},q,t)}{Z_{W_0}}=\sum_{K\geq 0}Q_{\star}^{K}\,\frac{W_{K}(Q_{i},q,t)}{Z_{W_0}}=1+Q_{\star}\frac{W_1}{Z_{W_0}}+\cdots\,,
\end{align}
with the expansion coefficients
\begin{align}
W_{K}(Q_{i},q,t)=\sum_{|\nu_{1}|+\cdots+|\nu_{k}|=K}Z_{\nu_{1}\cdots \nu_{k}}\,.
\end{align}
The function $Z_{\nu_{1}\cdots\nu_{k}}$ is the open string partition function which captures the contribution of open strings in the presence of Lagrangian branes with boundary conditions given by $\{\nu_{1},\cdots,\nu_{k}\}$ \cite{Aganagic:2000gs, Aganagic:2003db}. These Lagrangian branes are placed on the legs which get decompactified as $Q_{\star}\mapsto 0$. The function $W_{K}$ does not depend on $Q_{\star}$ and is well defined for any $K$. It captures the DT-invariants \cite{dt} for a class of sheaves supported on the curves labelled by the class of $Q_{\star}$. Since $W_{K}$ do not depend on $Q_{\star}$, in the limit $Q_{\star}\mapsto 0$ the partition function $Z_{W}$ reduces to $Z_{W_0}$.

\section{Intersection numbers for $X_{N,M}$}
\label{AppB}
In the following we give a very rough calculation of the intersection numbers in the elliptic Calabi-Yau threefold $X_{N,M}^{(\delta)}$, relying mostly on known results in the literature. 
\subsection{Infinite toric fan}
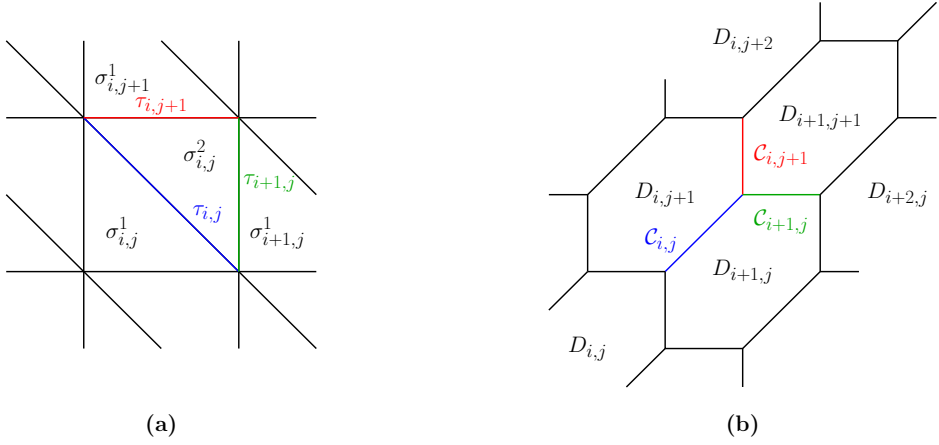
\begin{figure}[htb]
\begin{center}
\scalebox{0.34}{\parbox{37.5cm}{\begin{tikzpicture}[scale = 1.5]
\draw[ultra thick] (-2,0) -- (6,0);
\draw[ultra thick] (4,-2) -- (4,6);
\draw[ultra thick] (-2,4) -- (6,4);
\draw[ultra thick] (0,-2) -- (0,6);
\draw[ultra thick] (-2,6) -- (6,-2);
\draw[ultra thick] (2,6) -- (6,2);
\draw[ultra thick] (-2,2) -- (2,-2);
\draw[ultra thick, blue] (0,4) -- (4,0); 
\draw[ultra thick, red] (0,4) -- (4,4); 
\draw[ultra thick, green!65!black] (4,0) -- (4,4); 
\node at (3,3) {{\huge $\sigma_{i,j}^2$}};
\node at (1,1) {{\huge $\sigma_{i,j}^1$}};
\node at (5,1) {{\huge $\sigma_{i+1,j}^1$}};
\node at (1,5) {{\huge $\sigma_{i,j+1}^1$}};
\node at (3.2,1.5) {{\huge \color{blue} $\tau_{i,j}$}};
\node at (2,4.3) {{\huge \color{red} $\tau_{i,j+1}$}};
\node at (4.8,2.3) {{\huge \color{green!65!black} $\tau_{i+1,j}$}};
\node at (2,-4) {{\huge \textbf{(a)}}};
\begin{scope}[xshift=16cm,yshift=1cm]
\draw[ultra thick, blue] (-1,-1) -- (1,1);
\draw[ultra thick, red] (1,1) -- (1,3);
\draw[ultra thick, green!65!black] (1,1) -- (3,1);
\draw[ultra thick] (-1,-1) -- (-1,-3) -- (1,-3) -- (3,-1) -- (3,1);
\draw[ultra thick] (-1,-1) -- (-3,-1);
\draw[ultra thick] (-3,-1) -- (-3,1) -- (-1,3) -- (1,3);
\draw[ultra thick] (1,3) -- (3,5) -- (5,5) -- (5,3) -- (3,1);
\draw[ultra thick] (5,5) -- (6,6);
\draw[ultra thick] (5,3) -- (6,3);
\draw[ultra thick] (3,5) -- (3,6);
\draw[ultra thick] (-1,3) -- (-1,4);
\draw[ultra thick] (3,-1) -- (4,-1);
\draw[ultra thick] (1,-3) -- (1,-4);
\draw[ultra thick] (-3,1) -- (-4,1);
\draw[ultra thick] (-1,-3) -- (-2,-4);
\draw[ultra thick] (-3,-1) -- (-4,-2);
\node at (2,2) {{\huge \color{red} $\mathcal{C}_{i,j+1}$}};
\node at (2,0.3) {{\huge \color{green!65!black} $\mathcal{C}_{i+1,j}$}};
\node at (-1.1,-0.2) {{\huge \color{blue} $\mathcal{C}_{i,j}$}};
\node at (-1,1) {{\huge  $D_{i,j+1}$}};
\node at (1,-1) {{\huge  $D_{i+1,j}$}};
\node at (3,3) {{\huge  $D_{i+1,j+1}$}};
\node at (1,5) {{\huge  $D_{i,j+2}$}};
\node at (5,1) {{\huge  $D_{i+2,j}$}};
\node at (-3,-3) {{\huge  $D_{i,j}$}};
\node at (1,-5) {{\huge \textbf{(b)}}};
\end{scope}
\end{tikzpicture}}}
\end{center}
\caption{\sl (a) A local view of the $x-y$ plane at $z=1$ of an infinite toric fan with some maximal cones $\sigma$ and walls $\tau$ labeled. (b) A local view of the dual diagram to our infinite fan with some curves and divisors labeled.}
\label{Fig:WebDiagramGeneric}
\end{figure}

As in \cite{Kan}, we start by considering an infinite toric fan, which can be decomposed into the following set of maximal cones in $\mathbb{R}^3$:
\begin{align}
\sigma_{i,j}^1&= \mathbb{R}_{\geq 0}(i,j,1) + \mathbb{R}_{\geq 0}(i+1,j,1) +\mathbb{R}_{\geq 0}(i,j+1,1)\,, \nonumber \\
\sigma_{i,j}^2&= \mathbb{R}_{\geq 0}(i+1,j,1) + \mathbb{R}_{\geq 0}(i,j+1,1) +\mathbb{R}_{\geq 0}(i+1,j+1,1)  \,, \quad i,j\in \mathbb{Z}
\label{infiniteFan}
\end{align}
where the triples $(i,j,k)\in\mathbb{Z}^3$ are called ray generators in the following. Since all ray generators in (\ref{infiniteFan}) end on the same plane defined by $z=1$ in $\mathbb{R}$ (\emph{i.e.} $k=1$ in all cases), the resulting geometry is Calabi-Yau and non-compact. A local region of the toric fan looks as shown in \figref{Fig:WebDiagramGeneric}, where it is sufficient to show only the $x-y$ plane at $z=1$ due to the Calabi-Yau condition.
Each wall, that is the intersection of two maximal cones, defines an irreducible toric curve. There are three families of curves, diagonal (blue), horizontal (green), and vertical (red) (the orientations are defined with respect to the dual toric graph \figref{Fig:WebDiagramGeneric} (b)). The walls have the following form:
\begin{align}
\tau_{i,j}&= \sigma_{i,j}^{1} \cap \sigma_{i,j}^2 = \mathbb{R}_{\geq 0} (i+1,j,1) + \mathbb{R}_{\geq 0} (i,j+1,1)\,, \nonumber \\
\tau_{i+1,j}&= \sigma_{i,j}^{2} \cap \sigma_{i+1,j}^1 = \mathbb{R}_{\geq 0} (i+1,j,1) + \mathbb{R}_{\geq 0} (i+1,j+1,1)\,, \nonumber \\
\tau_{i,j+1}&= \sigma_{i,j}^{2} \cap \sigma_{i,j+1}^1 = \mathbb{R}_{\geq 0} (i,j+1,1) + \mathbb{R}_{\geq 0} (i+1,j+1,1)\,. \label{walls}
\end{align}
Due to $SL(2,\mathbb{Z})$ symmetry, the three families of curves are equivalent to each other, i.e. they are mapped into another in different $SL(2,\mathbb{Z})$ frames. It is thus sufficient to focus on one class of curves. We choose the diagonal (blue) one. In the following, we shall follow \cite{TV} for a general result (reduced to a three-dimensional fan) and apply it directly to the specific construction above: Let $u_{i=0,1,2,3}$ be four ray generators in a smooth three-dimensional toric fan. If $\tau= \sigma \cap \sigma'$ is a wall in the latter, which is defined through \begin{align}
&\tau= \mathbb{R}_{\geq 0}u_1 +\mathbb{R}_{\geq 0}u_2 \,,
&\sigma= \mathbb{R}_{\geq 0}u_0 + \mathbb{R}_{\geq 0}u_1 +\mathbb{R}_{\geq 0}u_2 \,,&& 
\sigma'= \mathbb{R}_{\geq 0}u_1+ \mathbb{R}_{\geq 0}u_2 +\mathbb{R}_{\geq 0}u_3\,,  \nonumber
\end{align}
there exist integers $b_{1,2}$ such that the wall relation  
\begin{align}
u_0 + b_1u_1+b_2u_2+u_3=0 \label{wrelation}
\end{align}
is satisfied. The intersection number of the irreducible curve $\mathcal{C}_{\tau}$ associated to $\tau$ with the divisor $D_{u}$ associated to any ray generator $u$ of the fan is then given by:
\begin{align}
D_u \cdot \mathcal{C}_{\tau}= \begin{cases}
1 \quad &\text{if } u=u_0,u_3\,, \\[-4pt]
b_i \quad &\text{if } u=u_i \, \text{ for } i=1,2\,, \\[-4pt]
0 \quad &\text{else}\,. \\
\end{cases}\label{IntersectionGeneral}
\end{align}
Specifically, for the toric fan (\ref{infiniteFan}), we have the wall relation for $\tau_{i,j}$ defined in (\ref{walls}):
\begin{align}
&(i,j,1) -1(i+1,j,1)-1(i,j+1,1)+(i+1,j+1,1)=0 \,,&&\text{with }&&b_1=b_2=-1\,.
\end{align}
Thus (\ref{IntersectionGeneral}) directly yields the following intersection numbers for the curves associated to $\tau_{i,j}$ in (\ref{walls}), with all divisors (associated with the ray generators $u$)
\begin{align}
D_u \cdot \mathcal{C}_{\tau_{i,j}}= \begin{cases}
1 \quad &\text{if } u=u_0,u_3 \\[-4pt]
-1 \quad &\text{if } u=u_k \, \text{ for } k=1,2 \\[-4pt]
0 \quad &\text{else} \\
\end{cases}
\end{align}
By $SL(2,\mathbb{Z)}$ symmetry, we thus have the following non-zero intersection numbers for the dual toric diagram in \figref{Fig:WebDiagramGeneric} (b):
\begin{align}
D \cdot \mathcal{C}_{i,j}&= \begin{cases}
1 \quad &\text{if } D=D_{i,j},D_{i+1,j+1}\,, \\[-4pt]
-1 \quad &\text{if } D=D_{i+1,j},D_{i,j+1}\,, \\
\end{cases}
\quad \quad
D \cdot \mathcal{C}_{i+1,j}= \begin{cases}
1 \quad &\text{if } D=D_{i,j+1},D_{i+2,j}\,, \\[-4pt]
-1 \quad &\text{if } D=D_{i+1,j},D_{i+1,j+1}\,, \\
\end{cases} \nonumber \\
D \cdot \mathcal{C}_{i,j+1}&= \begin{cases}
1 \quad &\text{if } D=D_{i+1,j},D_{i,j+2}\,, \\[-4pt]
-1 \quad &\text{if } D=D_{i,j+1},D_{i+1,j+1}\,. \\
\end{cases}
\end{align}
Summarizing these results in words, we can say: 
The intersection of a curve $\mathcal{C}$ with a divisor $D$ is $1$ if $\mathcal{C}$ sticks out $D$ and it is $-1$ if $\mathcal{C}$ lies inside $D$.
\subsection{$X_{N,M}^{(\delta)}$ and intersection numbers}
\begin{wrapfigure}{R}{0.4\textwidth}
\begin{center}
\scalebox{0.4}{\parbox{10cm}{\begin{tikzpicture}[scale = 1]
\draw[ultra thick] (-1,-1) -- (1,1);
\draw[ultra thick] (1,1) -- (1,3);
\draw[ultra thick, red] (1,1) -- (3,1);
\draw[ultra thick] (-1,-1) -- (-1,-3) -- (1,-3) -- (3,-1) -- (3,1);
\draw[ultra thick] (-1,-1) -- (-3,-1);
\draw[ultra thick] (-3,-1) -- (-3,1) -- (-1,3) -- (1,3);
\draw[ultra thick] (1,3) -- (3,5) -- (5,5) -- (5,3) -- (3,1);
\draw[ultra thick] (5,5) -- (6,6);
\draw[ultra thick] (5,3) -- (6,3);
\draw[ultra thick] (3,5) -- (3,6);
\draw[ultra thick] (-1,3) -- (-1,4);
\draw[ultra thick] (3,-1) -- (4,-1);
\draw[ultra thick] (1,-3) -- (1,-4);
\draw[ultra thick] (-3,1) -- (-4,1);
\draw[ultra thick] (-1,-3) -- (-2,-4);
\draw[ultra thick] (-3,-1) -- (-4,-2);
\node at (2,0.3) {{\huge \color{red} $\mathcal{C}$}};
\node at (1,-1) {{\huge  $D_1$}};
\node at (3,3) {{\huge  $D_2$}};
\node at (-1,1) {{\huge $D_3$}};
\end{tikzpicture}}}
\caption{\sl A curve $\mathcal{C}$ and divisors $D_1$, $D_2$ and $D_3$.}
\end{center}
\label{Int1}
\end{wrapfigure}
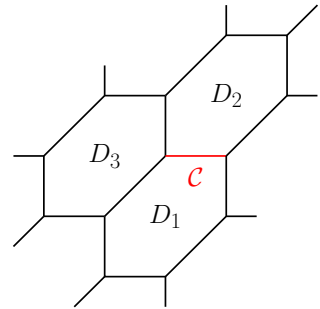
In \cite{Kan}, the authors gave a toric construction of $X_{N,M}$. Roughly speaking, they consider an infinite toric fan quotiented by $N \mathbb{Z} \times M \mathbb{Z}$ to impose periodic identifications in the web diagram. In the following we assume\footnote{Although we have not checked explicitly the existence of a quotient that satisfies all consistency conditions given in \cite{Kan}, we have checked our results for the intersection numbers (\ref{inter1}) and (\ref{inter2}) in various cases through other methods. In particular, for the cases $X^{(\delta)}_{2,1}$, $X_{2,2}^{(\delta)}$ (for $\delta=0,1$) and $X_{2,N}^{(\delta)}$ (for $\delta=0,1,\ldots,N-1$) we have calculated them independently by representing the geometry locally as (combinations of) $\mathbb{P}^1\times \mathbb{P}^1$ and found complete agreement. This leads us to believe that a quotient procedure as detailed below can be employed to compute the intersection numbers.}
that there exists a similar quotient, which gives rise to the periodic identifications required in the web diagram $X_{N,M}^{(\delta)}$. For most curves and divisors, this quotient does no change the intersection numbers as devised in the previous section. Nevertheless, in the webs $X_{N,M}^{(\delta)}$ with $N=1$ or $M=1$ some curves will see their intersections numbers modified due to the fact that a given curve $\mathcal{C}$ can now intersect two irreducible divisors $D$ and $D'$ that are identified under the quotient procedure. We will discuss in the following the two special situations that may arise for $N$ or $M$ equal to 1, thus changing the effective rule given at the end of the previous section. It is sufficient to focus on configurations of type $X_{N,1}$ as the $X_{1,M}$ configuration will follow the same pattern by $SL(2,\mathbb{Z})$ symmetry.\\[-8pt]

\noindent 
$\bullet$ {\bf Case 1:} The curve $\mathcal{C}$ lies inside two divisors which get identified under the quotient action. In the infinite fan we can consider the intersection of the toric curve $\mathcal{C}$ with the divisor $D=D_1+D_2$ in \figref{Int1}:
\begin{align}
\mathcal{C}\cdot D=\mathcal{C} \cdot D_1 + \mathcal{C} \cdot D_2=1+1=2 \label{inter1}
\end{align}
Under the quotient action, the two irreducible divisors get identified $D_1 \sim D_2 $, leading to the result above. \\

\noindent
$\bullet$ {\bf Case 2:} The curve $\mathcal{C}$ lies in one divisor and sticks out of another one and both get identified under the quotient. In terms of the infinite fan we are interested in the intersection of $\mathcal{C}$ with the divisor $D'=D_1+D_3$ (see \figref{Int1})
\begin{align}
\mathcal{C}\cdot D'=\mathcal{C} \cdot D_1 + \mathcal{C} \cdot D_3=1-1=0\,. \label{inter2}
\end{align}


\end{document}